\PassOptionsToPackage{table,xcdraw}{xcolor}
\documentclass[format=acmsmall, review=false, screen=true]{acmart}

\usepackage[ruled]{algorithm2e} 
\usepackage{graphicx} 
\usepackage{xcolor,colortbl}
\usepackage{multirow}
\usepackage{bbding}
\usepackage{pifont}
\usepackage{wasysym}
\usepackage{amssymb}
\usepackage{caption}
\usepackage{xcolor,colortbl}
\usepackage{booktabs} 

\usepackage{longtable}
\usepackage{afterpage}  
\usepackage{tabularx} 

\usepackage{amssymb}
\usepackage{pifont}
\newcommand{\xmark}{\ding{55}}%

\usepackage{lscape}
\usepackage{rotating}
\usepackage{sidecap}
\sidecaptionvpos{figure}{c}

 \newcolumntype{b}{>{\hsize=2.3\hsize}X}
 \newcolumntype{m}{>{\hsize=.9\hsize}X}
 
\acmVolume{9}
\acmNumber{9}
\acmArticle{99}
\acmYear{9999}
\acmMonth{9}

\setcopyright{acmcopyright}

\acmDOI{0000001.0000001}

\begin{document}
	
	\title{Designing Security and Privacy Requirements in  Internet of Things: A Survey}
	
	
	\author{Nada Alhirabi}
	\orcid{1234-5678-9012-3456}
	\affiliation{%
		\institution{Cardiff University}
		\streetaddress{School of Computer Science and Informatics}
		\city{Cardiff}
		\postcode{CF24 3AA}
		\country{UK}}
	\email{alhirabin@cardiff.ac.uk}
	
	\author{Omer Rana}
	\affiliation{%
		\institution{Cardiff University}
		\streetaddress{School of Computer Science and Informatics}
		\city{Cardiff}
		\postcode{CF24 3AA}
		\country{UK}}
	\email{ranaof@cardiff.ac.uk}

	\author{Charith Perera}
	\affiliation{%
		\institution{Cardiff University}
		\streetaddress{School of Computer Science and Informatics}
		\city{Cardiff}
		\postcode{CF24 3AA}
		\country{UK}}
	\email{pererac@cardiff.ac.uk}

	\renewcommand\shortauthors{Alhirabi, N. et al}
	
	\begin{abstract}
			
		The design and development process for the Internet of Things (IoT) applications is more complicated than that for desktop, mobile, or web applications. First, IoT applications require both software and hardware to work together across many different types of nodes with different capabilities under different conditions. Secondly, IoT application development involves different types of software engineers such as desktop, web, embedded and mobile to work together. Further, non-software engineering personal such as business analysts are also involved in the design process. In the addition to the complexity of having multiple software engineering specialists cooperating to merge different hardware and software component together, the development process required different software and hardware stacks to integrated together (e.g., different stacks from different companies such as Microsoft Azure, IBM Bluemix). Due to above complexities, more often non-functional requirements (such as security and privacy which are highly important in the context of IoT) tend to get ignored or treated as second class citizens in IoT application development process. In this paper, we have reviewed techniques, methods and tools that are being developed to support incorporating security and privacy requirements into traditional application designs.  By doing so, we aim to explore how those techniques could be applicable to the IoT domain.
		During our review, we realised that by far the majority of the requirement engineering efforts are focused on security. In this paper, we primarily focused on two different aspects: (1) design notations, models, and languages that facilitate capturing non-functional requirements (i.e., security and privacy), and (2) proactive and reactive interaction techniques that can be used to support and augment the IoT application design process. Our goal is not only to analyse, compare and consolidate past research work but also to appreciate their findings and discuss their applicability towards the IoT.
		

		
	\end{abstract}

	%
	%

	
	%
	%


	\maketitle

	\section{Introduction}
	
	Until 2003, Cisco IBSG's \cite{evans2011internet} did not recognise IoT due to the few numbers of connected devices where there was only 0.08 device per person. With the exponential growth of smartphones, tablets, smart devices and their applications over the years, connected things significantly increase for each person. In 2010, there were 6.8 billion people using 12.5 billion devices which are 1.84 devices per person. Due to technological enhancements in distributed affordable sensors, contactless data exchange such as RFID, short-range wireless such as Bluetooth and ZigBee, and internet mobile access, a global network of connected things has increased \cite{atzori2010internet}. It is predicted to reach 20-24 billion connected devices by 2020 \cite{VanderMeulen2017} \cite{gubbi2013internet}.
	
	Designing and developing IoT applications is complicated compared to the one created for desktop, web, or mobile applications. 
	First, IoT applications need to support different technologies working together such as hardware/firmware, software, sensor, semantic, cloud, data storing, data modelling, processing, and communication technologies \cite{perera2014context}. All of these components are working throughout many different types of nodes under different conditions and challenges, and all of them could be vulnerable to attacks \cite{gubbi2013internet}. Moreover, applications should consider many features that IoT needs to support such as devices heterogeneity, scalability, ubiquitous data exchange, semantic interoperability and data management, etc. \cite{MIORANDI20121497} \cite{chatzigiannakis200750}. In order to the Internet of Things to be successful, the development process needs to consider the whole connecting devices, and improving the scenarios used in traditional mobile computing \cite{gubbi2013internet}.
	
	In addition to IoT heterogeneity nature, IoT applications involve the cooperation of multiple software engineers whose having different expertise. Those engineers are working together on different components with different application domains such as home automation, smart cities, smart driving...etc. Due to the lacking of full-stack developers, the development process needs to have a common programming framework to support developers needs \cite{giang2015developing}. Moreover, end-users preferences are divers and they need to create these preferences using graphical user interface (GUI) at run-time. Consequently, applications developers and devices manufacturers should focus on the end-user requirements \cite{heo2015iot}.

	As a result of the stated complexities, non-functional requirements (NFRs), such as security and privacy, have not received sufficient attention \cite{Trujillo2009} especially in the traditional Software Development Lifecycle (SDLC). It has been stated that the main source of software vulnerabilities is discovered to be in the early stages of the SDLC and the majority of them could be eliminated at this step \cite{geer2010companies}. Consequently, adopting security and privacy in SDLC is becoming critical, and embedding them in the early stage of the SDLC become an insistent \cite{tondel2008security} \cite{geer2010companies}. Microsoft took a step forward and introduced Microsoft's Security Development Lifecycle (SDL) that consists of practices for supporting security \cite{Microsofta}. 
	
	With the raising connectivity towards the IoT and having massive of new devices and applications that are connected to the internet every year, threats keep changing \cite{ howard2006security}. The burden of the vulnerabilities in the traditional SDLC become bigger with the IoT heterogeneity. Security by Design (SbD) has been approved that it is an effective way to create a secure system by emerging security at an early stage of the software development lifecycle (SDLC). As well as SbD, Privacy by Design (PbD) is recently recognized to be an important and useful approach for preserving privacy for software-based systems \cite{Sion2018b}. This importance is confirmed by REGULATION (EU) 2016$/$679 General Data Protection Regulation (GDPR) \cite{EuropeanUnion2016} which is applied to all systems that deal with personal data processing which is common in IoT applications. The contribution of this paper is as follows:
	\begin{itemize}
		\item Review the evolution of design notations, models, and languages that facilitates capturing non-functional requirements (i.e., security and privacy).
		\item Propose and use a taxonomy to compare and contrast past approaches.
		\item Review a few selected tools that have been built to facilitate capturing non-functional requirements from two different perspectives: (i) proactive and (ii) reactive interaction techniques.
	\end{itemize}
	
	
	
	
	\subsubsection*{Paper structure} The  paper is structured into sections as follows: section \ref {Section:Traditional} presents background information about the traditional software development life cycle and its phases.  Section \ref{Section:OverviewIoT} gives a brief IoT overview and how IoT differ from embedded systems. This section includes a case scenario that will be used in a later section. Section \ref{Section:IotDLC} explains the difference between the distributed nature of IoT application development and traditional SDLC. In section \ref{Section:FRs_NFRS}, we briefly introduce functional and non-functional requirements. Literature search and selection methodology are introduced in section \ref{Section:Methodology}, with some details about search queries and steps of data selection and extraction. Section  \ref{Section:RESULTS_OF_THE_REVIEW} and \ref{Section:DesignPrinciples}  gives an exploratory analysis of some of the available design notations and design principle respectively. Section  \ref{Section:ResearchChallenges} discusses the survey limitations. Section \ref{Conclusion} concludes the survey.

	\section{Traditional Software Development Life Cycle}\label{Section:Traditional}
	Let us now briefly introduce software development life cycle (SDLC) in general as  it helps to understand the remaining sections of the paper. In later sections, we will  discuss specific challenges in IoT and their impact on SDLC.
	
\begin{figure}[b!]
		\vspace{-12pt}
		\includegraphics[scale=0.5]{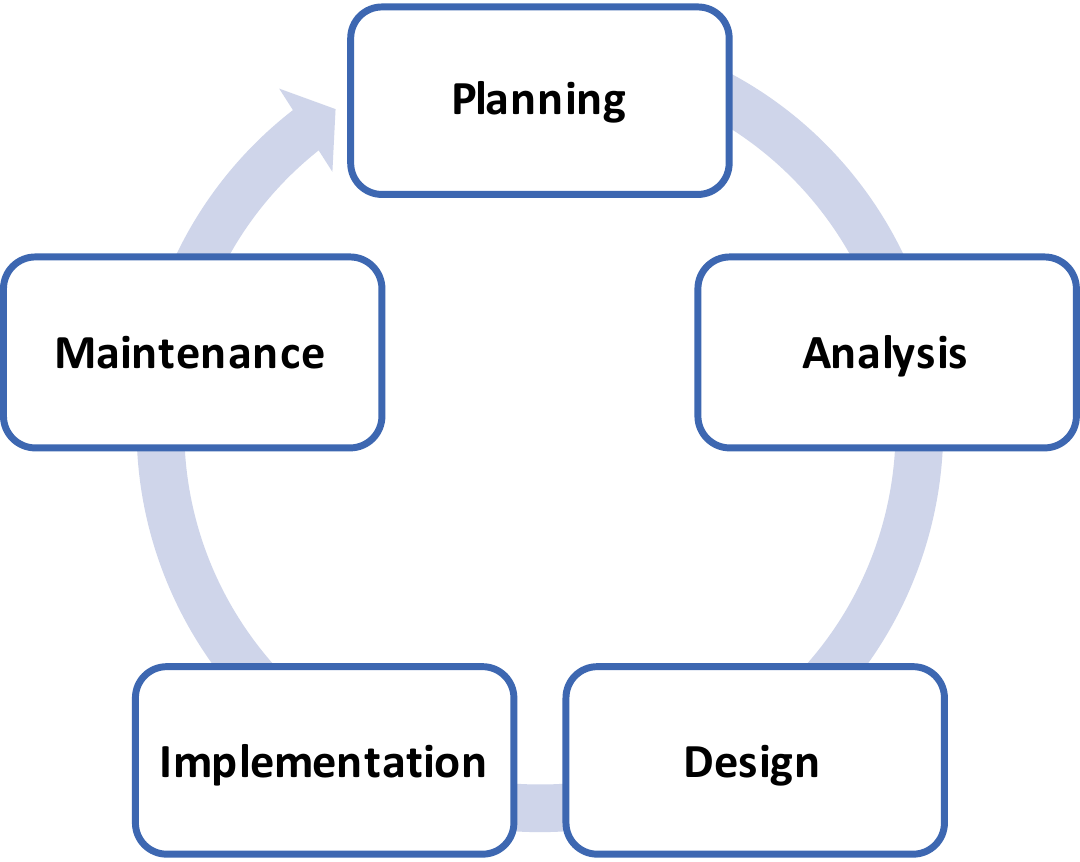}
		\caption{Software Development Life Cycle (SDLC) based on \cite{ hoffer2012modern} and \cite{Valacich}}
		
		\label{SDLC}
	\end{figure}

	In software engineering, SDLC is the most significant element. SDLC is a traditional methodology (or process) used for building and maintaining software systems where some essential phases should be followed. In general, software developments models have three primary goals which are improving the system quality, providing management controls and maximizing productivity. There are several different software development life cycle models. Each one is developed for specific purposes. According to Hoffer \cite{ hoffer2012modern} and Valacich \cite{Valacich}, SDLC consists of five phases which are planning, analysis, design, implementation and maintenance as shown in Figure \ref{SDLC}. 
	
	The planning phase is used for identifying and analysing information system needs. Then, these needs are prioritized and translated into a development schedule. After that, the proposed problem is investigated to determine the proposed system scope. In analysis phase, requirements are determined, and some alternative designs are suggested and compared.  After the requirements are analysed, they become well-defined and documented. The outcome of analysis phase is software requirement specification (SRS) document. SRS document lists all the essential requirements that needed to be designed and developed. At the end of this stage, SRS document is handled to the end-user to approve it \cite{Models2013}.
	
	In the design phase, all the provided description will be converted into logical and then physical system specifications. The logical design is independent of any software, hardware or platform. It focusses on business features of the system. While in physical design, the specifications are transformed into technical specifications. In the implementation phase, six major activities are done which are coding, testing, installation, documentation, training and support. The goal of this phase is to translate the physical system specifications into reliable working software. It also record all the work that has been done. This phase also is considering the support for all users current and future once \cite{hoffer2012modern}. The last phase is maintenance phase, where fixing any issues founded by the customer/end-user is performed to keep the system working well.  Although the previous described SDLC gives a general overview of the systems development process, there are very specific methods that use the idea of the SDLC with some additions. For example, Microsoft's Security Development Lifecycle  (SDL) is a specialized SDLC that consists of practices for supporting security \cite{Microsofta}.

	As mentioned earlier, system development has many models such as waterfall model, spiral, v model, prototype, agile model etc. Each model has its own features, drawback and usage. Initially traditional development was introduced with some extends such as in waterfall, v model, incremental and spiral models. In any model of them, each phase of SDLC has deliverables and outcomes that used for the next phase in the life cycle.  They differ about the way that life cycle is arranged and executed as seen in Table \ref{SDLRep}.  There are some weaknesses related to these types of models, such as high effort, time and cost to change after a milestone is delivered. 
	
	As known, all the models, regardless of their names or types, have somewhat of requirement management and design phase. Either way it is only done at the beginning of the model or repeated as an iteration through the system development. There are many consequences when managing requirements and design are only done once at early stage of the system development. The major effect is the high-cost of restarting or remodelling the system once major mistakes happened. The other impact that gradually affect the cost happens when end-user requirements are keep changing while developing the system where changes in this stage are expensive.
	As stated by Geer \cite{geer2010companies}, the main source of software vulnerabilities is exposed in the early stage of the SDLC and the most of them could be excluded at this step. 
	Consequently, it is important factor for cost reduction to highlight any vulnerabilities/mistakes at design stage which is much cheaper than fixing them in later stages.

	As a result, other types of modelling such as agile modelling, as illustrated in Table \ref{SDLRep}, and object-oriented analysis and design (OOAD) are introduced to overcome the traditional SDLC limitations where remodelling/restarting is not expensive. A short comparison between traditional development, agile development and object-oriented analysis and design are given in Table \ref{SDL}. This comparison is based on some criterias such as primary objectives, requirement, cost of restarting and remodeling, etc. This table can give to some extent an overview about gathering and fulfilling user requirements and the cost of change in each model. It could help system engineers to clearly decide which model to follow before developing any system.
	
	\begin{table}\sffamily 
		\footnotesize 
		\begin{tabular}{| p{6.5cm}|| p{6.5cm}| }
			\toprule
			
			\multicolumn{2}{|c|}{List of graphical illustration for some Software Development Models } \\ 
			\midrule
			\includegraphics[scale=.285]{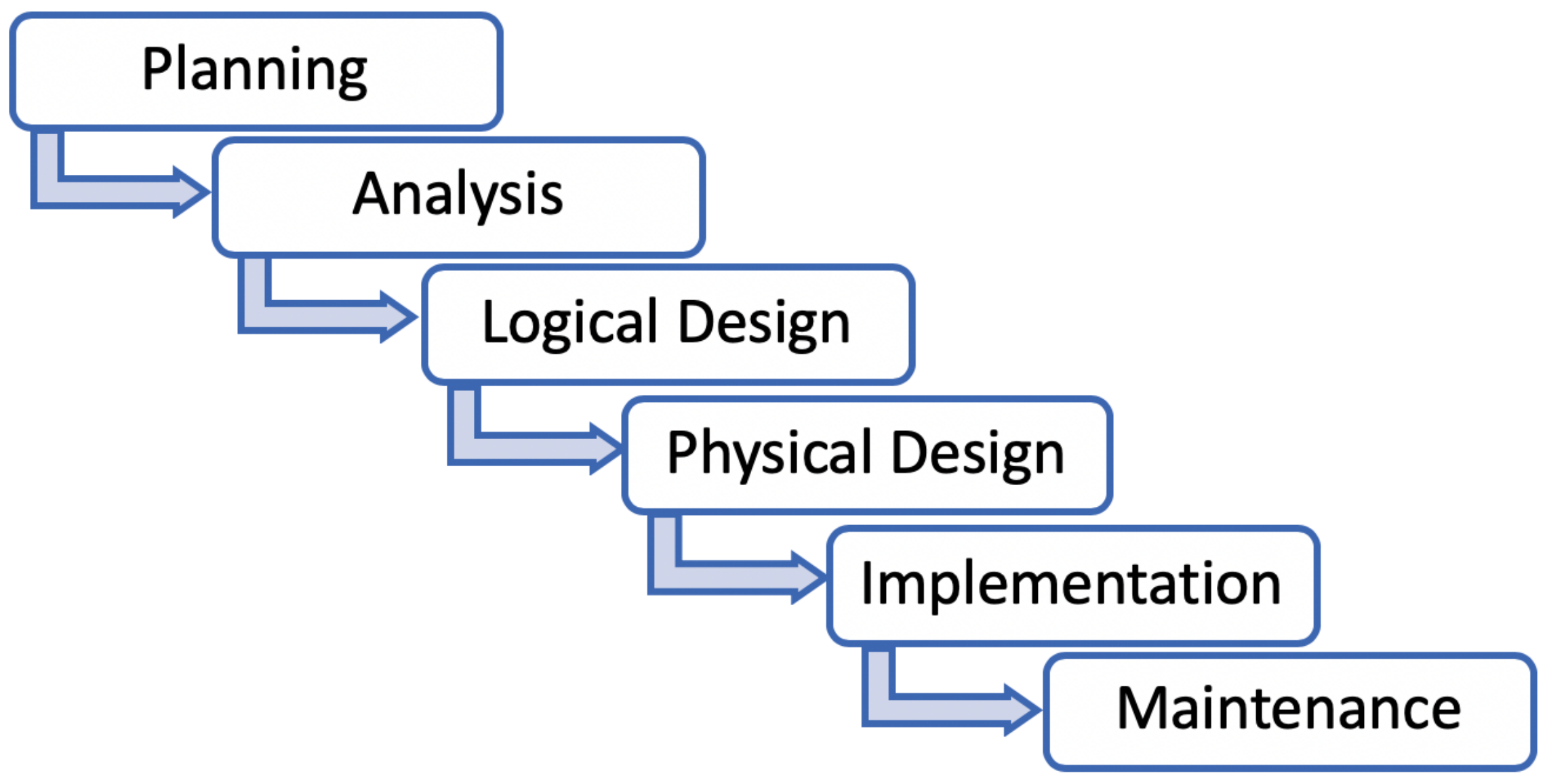} &\includegraphics[scale=.25]{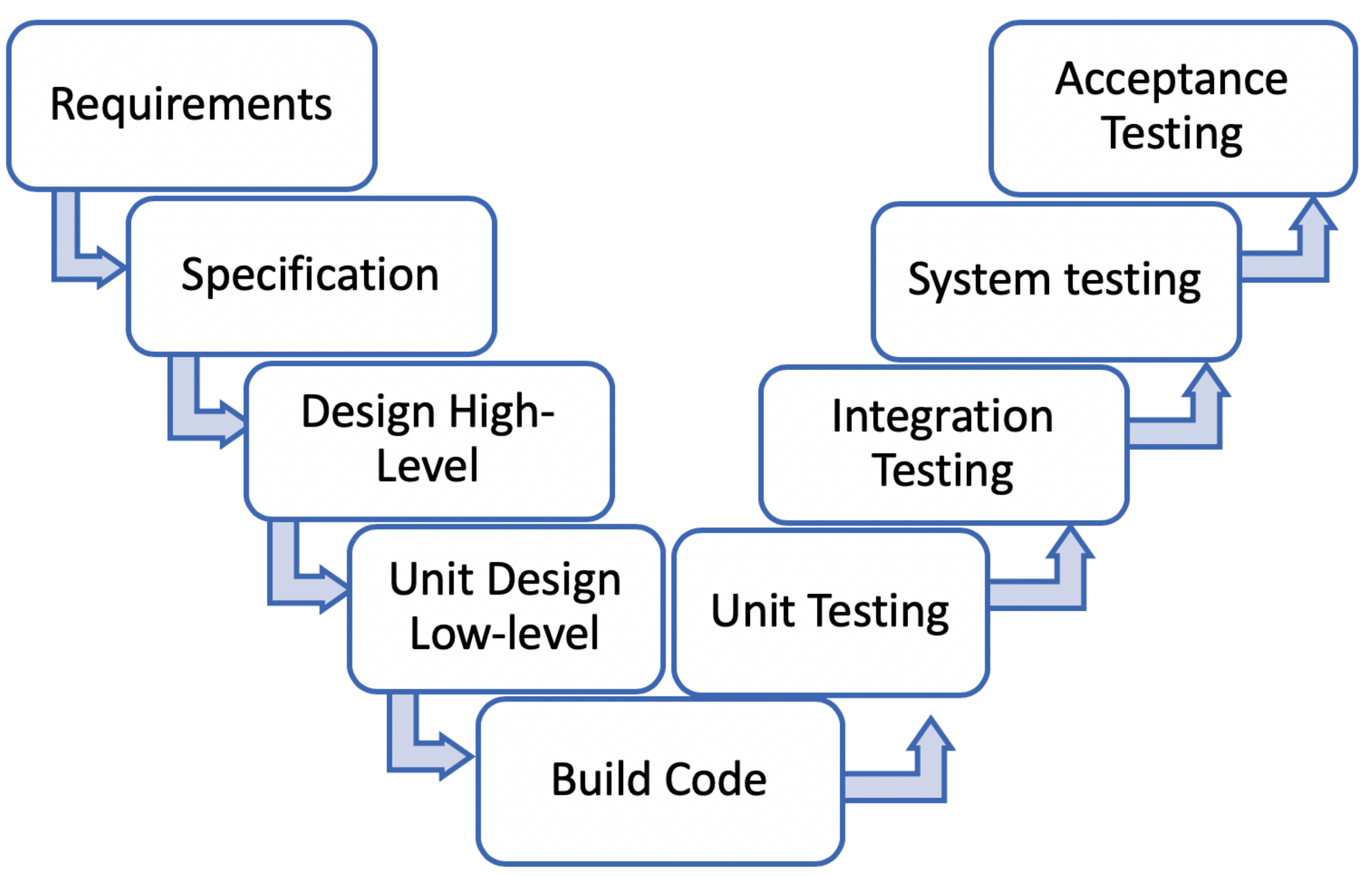} \\ 
			
			Waterfall model. \textit{Based on} \cite{Valacich}.  	&V model. \textit{Based on} \cite{Balaji2012}. 		 	 \\\\
			\midrule

			\includegraphics[scale=.27]{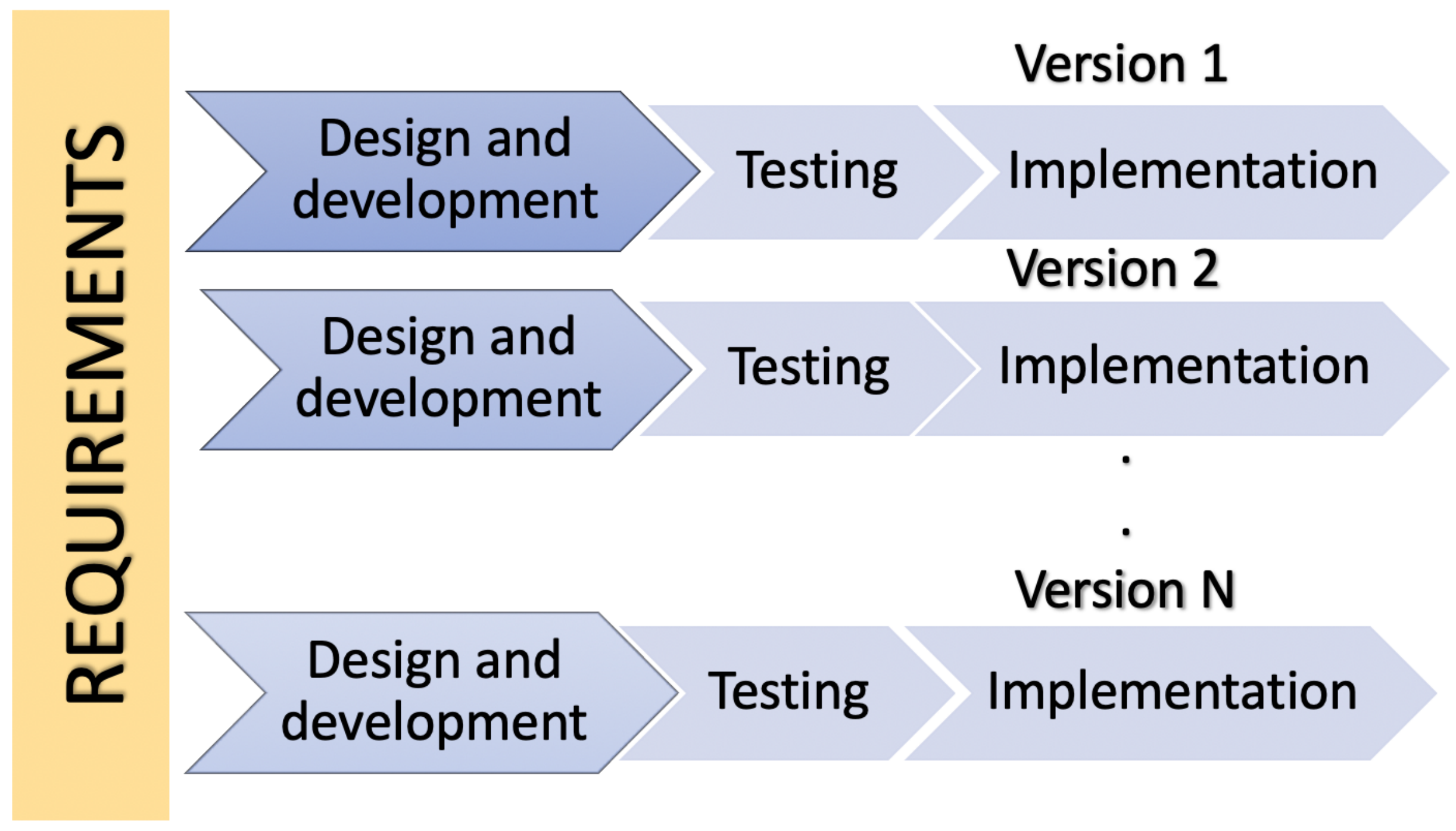} &\includegraphics[scale=.13]{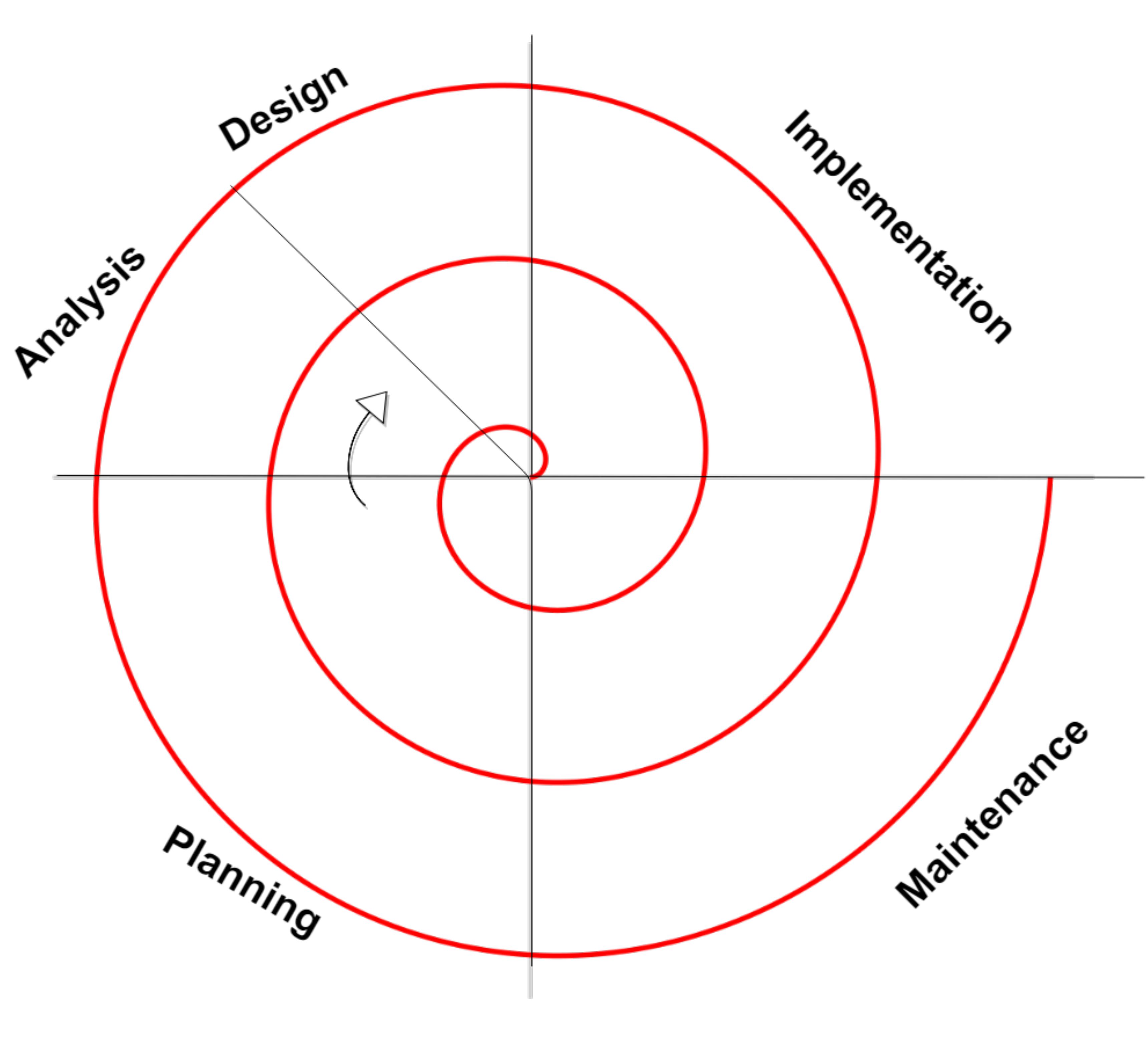} \\ 
			Incremental model. \textit{Based on} \cite{Models2013}. & Spiral model. \textit{Based on} \cite{Valacich}. \\\\	
			
			\midrule
			\includegraphics[scale=.24]{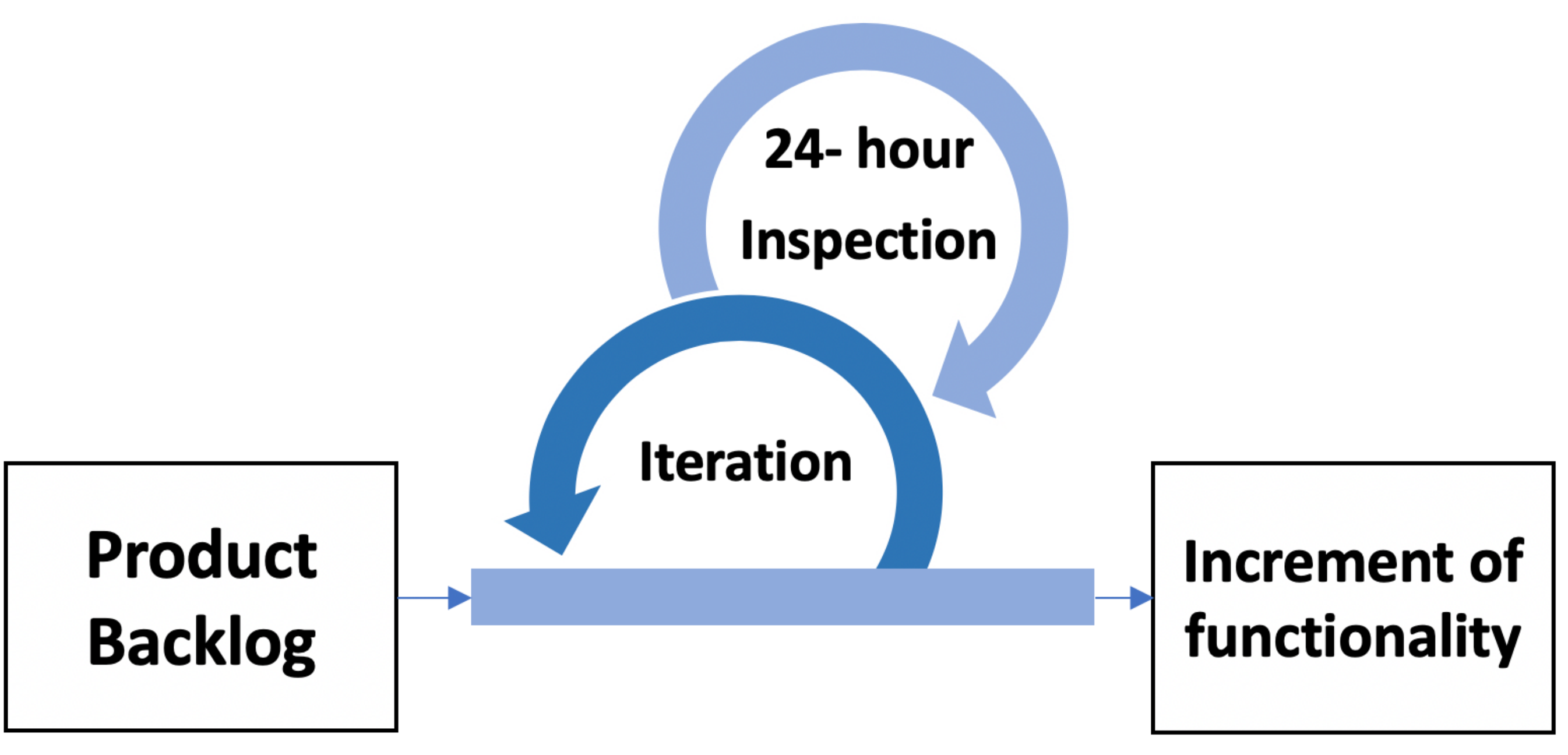} &  \includegraphics[scale=.29]{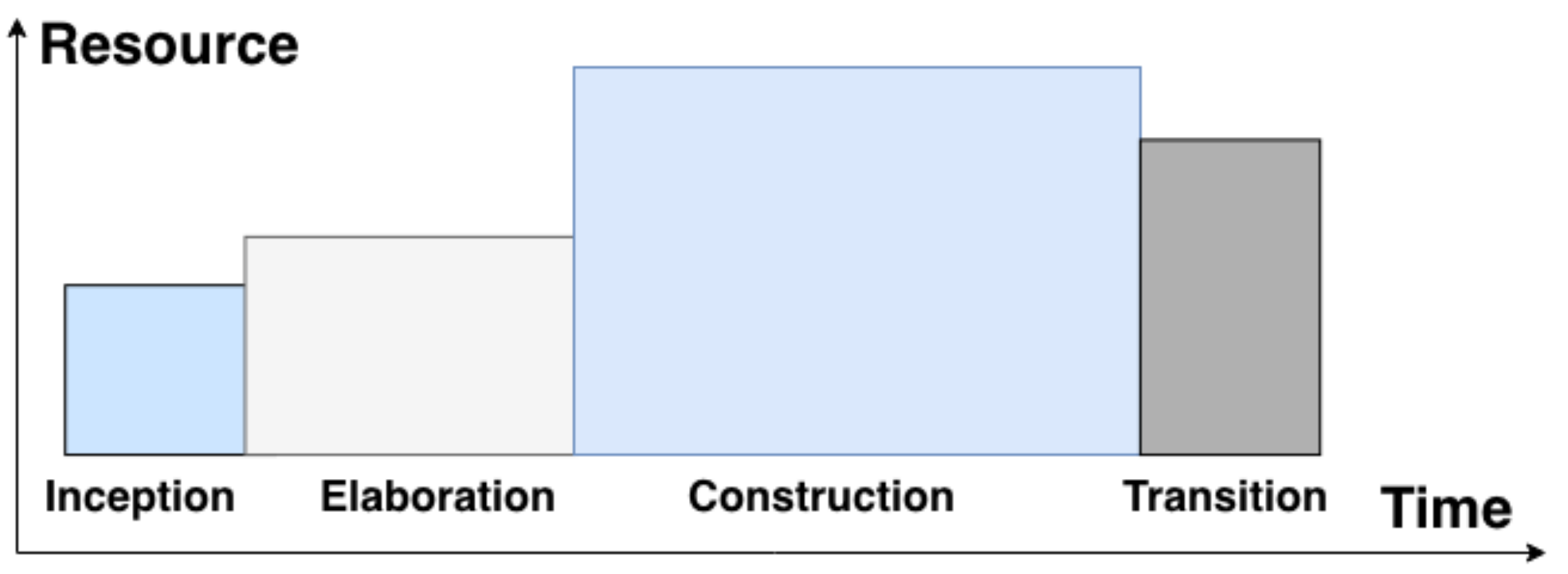}\\
			Scrum (Agile model). \textit{Based on} \cite{schwaber2004agile}.	 	& 	Rational Unified Process (RUP). \textit{Based on}	\cite{Valacich}.\\	
			
			\bottomrule 
		\end{tabular}
		\caption{List of graphical illustration for some Software Development Models: Waterfall model, V model, Incremental model, Spiral model, Scrum (Agile model) and Rational Unified Process (RUP) model.  All of them  vary on the way the life cycle is arranged and applied.}
		\label{SDLRep}
		\vspace{-22pt}
	\end{table}

	
	{\footnotesize
		\begin{table}
			\begin{tabular}{ |p{3cm}||p{3cm}|p{3cm}|p{3cm}|  }
				\hline
				\multicolumn{4}{|c|}{Software Development Models List} \\
				\midrule
				\cellcolor{gray!25}		&Traditional development 	&Agile development	&object-oriented analysis and design (OOAD)\\
				\midrule\midrule
				Development model  	& Life cycle model.    &Evolutionary-delivery model.&   Object-Oriented approach.\\
				\midrule
				Primary objectives		& Safety. Too many  process and documentation is done for safety wich lead to slow development.  & Quick results due to doing many iterations.		&Quality and productivity by focussing in inheritance for refutability.\\
				\midrule
				Organizational structure			&   Formal, targeting large organizations.  & Flexible, targeting small\textbackslash medium organizations.   &Flexible, focussing on objects.\\
				\midrule
				User requirements   		&Well-defined before
				implementation. & Co-operative input.& Co-operative. \\
				\midrule
				Changes adaption		&   Poor changes adaption. Models such as waterfall does not allow changes of defined requirements as the project is progressing. 
				
				&Changes and corrections are one each iteration.   &Changes and corrections are one each iteration.  \\			\midrule 
				
				
				End-product fulfilling the requirements
				& Since changes are limited through the development, it is potential that the software could not fully meet the end-user requirement. & Since user is co-operating and changes are repeatedly applied, it is potential that the software would meet the end-user requirement. & Since user is co-operating and changes are repeatedly applied, it is potential that the software would meet the end-user requirement.
				\\			\midrule

				Cost of restarting		&   High.  & Low. & cost is reduced. For each  iteration, full assessment is done for needed correction \cite{Hirsch2002}.\\
				\midrule
				Cost of remodeling		&   Expensive.  &  Not expensive due to user engagment and repeated assessment. & Not expensive due to continuous assessment.\\
				\midrule
				Testing	& Done after coding.  & Done each iteration.   & Done in iterations. \\
				\midrule
				Developers		& Organized with a plan.  & Co-located and
				interactive.	&Interactive.\\
				\midrule

				
				Size of  teams and projects		& Large.  & Small.	&Large-scale projects. There are some extends to OOAD approaches, such as Rational Unified Process (RUP) \cite{Monteiro:2012:RSR:2664360.2664387}, to support small\textbackslash micro software teams\textbackslash projects. \\
				\midrule
			\end{tabular}
			
			\caption{Differences between traditional, agile  and object-oriented analysis and design (OOAD) development approach. OOAD shares the iterative method from agile model, consequently they share some characteristics \cite{Models2013}\cite{Valacich}\cite{Hirsch2002}}.
			\label{SDL}
			\vspace{-20pt}
		\end{table}
	}


	\section{Overview of Internet of Things}\label{Section:OverviewIoT}
	Let us now briefly introduce IoT applications and their characteristics in general. In subsequent sections, we discuss SDLC from IoT perspective highlighting unique aspects and challenges of IoT. According to Gartner, the Internet of Things is the network of physical objects that contain embedded technology to communicate and sense or interact with their internal states or the external environment \cite{VanderMeulen2017}. These physical objects or things can be buildings, devices, automobiles, and other objects that embedded with sensors, software, electronics and network connection. DNA analysis devices, implanted heart monitoring and biochip transponders on animals are some examples of things that support the IoT concept. These devices gather suitable data using a variety of existing technologies and then move these data between other devices.
	
	\vspace*{-2mm}
	\paragraph{IoT history}
	Since the internet is founded in 1989, the idea of connecting things on the internet has been started. However, the term of Internet of Things (IoT) did not become popular till 1999 when it is created by Kevin Ashton, executive director of the Auto-ID Centre, MIT. At the same year, a global Radio-frequency identification (RFID) was invented which considered as a starting point for IoT. In 2000, LG company revealed its plan to produce a smart refrigerator that would determine by itself whether the food items are filled or not.
	The major trigger of IoT was in 2011 when IPv6 was lunched. Since
	then many famous IT companies such as Cisco, Ericson and IBM started to initiate many of IoT educational and commercial applications \cite{suresh2014state}.

	\subsection{{Difference Between IoT and Embedded Systems}}
	
\begin{figure}
		\includegraphics[scale=.4]{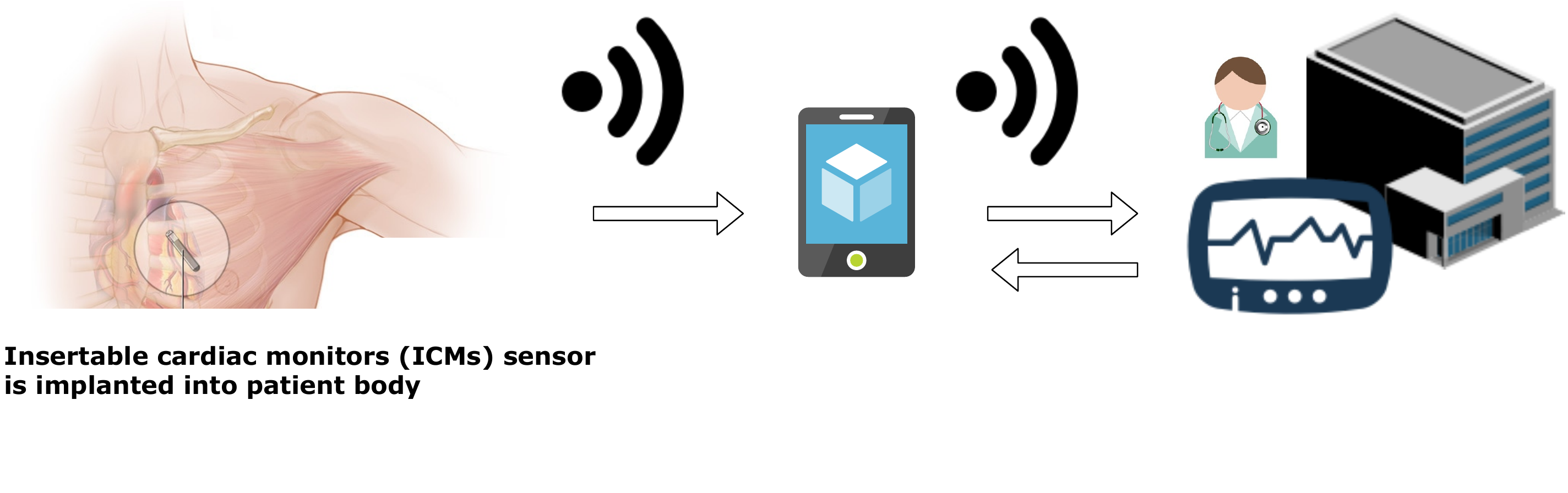}
		\caption{Insertable cardiac monitors (ICMs)}
		\label{Heart}
		\vspace*{-5mm}
	\end{figure}

	Embedded systems, called sometimes embedded chipsets, are computing devices that are sitting on the edge of an IoT product.  Connecting sensors to the internet is the main responsibility of the embedded system. It is also managing the communication between sensors and network by gathering, packaging and sending the data to an application. Furthermore, it may execute local application and security.
	Embedded system usually consists of a microcontroller that programmed to do a specific job. Once an embedded device is given access to the internet, it becomes an IoT device. For example, a lower power-consuming device such as implanted heart monitoring, showed in Figure \ref{Heart}, considered as an embedded device. It becomes an IoT device if it can transmit signal and pulses reading to a server's log via the Internet. These readings are checked then to see if there is an abnormality level. If there is, the server sends an alert to the concerned patient,s family mobile and doctors with some details such as the condition of the patient.
	
\begin{figure}
		\includegraphics[scale=.43]{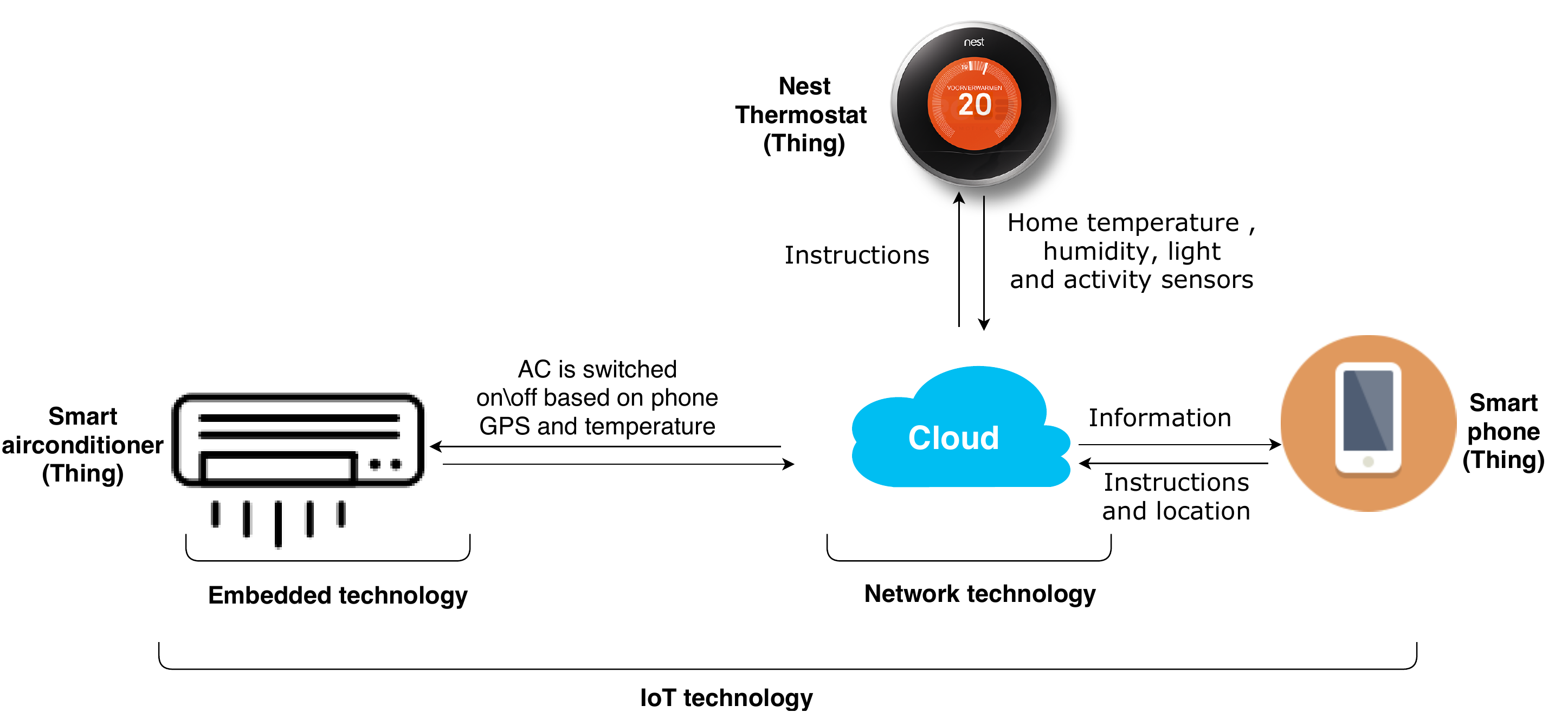}
		\caption{IoT Vs. Embedded systems}
		\label{EmbvsIoT}
		\vspace*{-5mm}
	\end{figure}

	Another example that can illustrate the idea of embedded system and IoT is smart air conditioner (AC) and Nest learning thermostat as seen in Figure \ref{EmbvsIoT}.  Nest thermostat is a programmable smart thermostat with self-learning Wi-Fi developed by Nest Labs. It can conserve energy by adjusting the heating and cooling of homes and businesses. Nest has an embedded system that sends home temperature, humidity, light and activity sensors. The smart AC by itself has an embedded system that can switch the AC on\textbackslash off based on the temperature sensor.  It becomes an IoT system once AC is connected to the internet, the cloud on this example, and interacted with the Nest temperature device and app. By taking advantage of Nest and AC sensors and phones' locations, the energy can be turned to saving mode when nobody is at home.

	\vspace*{-2 mm}
	
	\subsection{\textbf{Example IoT scenario}}
	In this section, diabetes treatment and monitoring use case scenario is presented from a problem owner's perspective. Developing an IoT application can solve this problem. This scenario highlights some privacy challenges as we will explain later.
	
	\subsubsection*{Use case: diabetes treatment and monitoring}\label{sec:CGM_uscase}
\begin{figure}
		\includegraphics[scale=0.35]{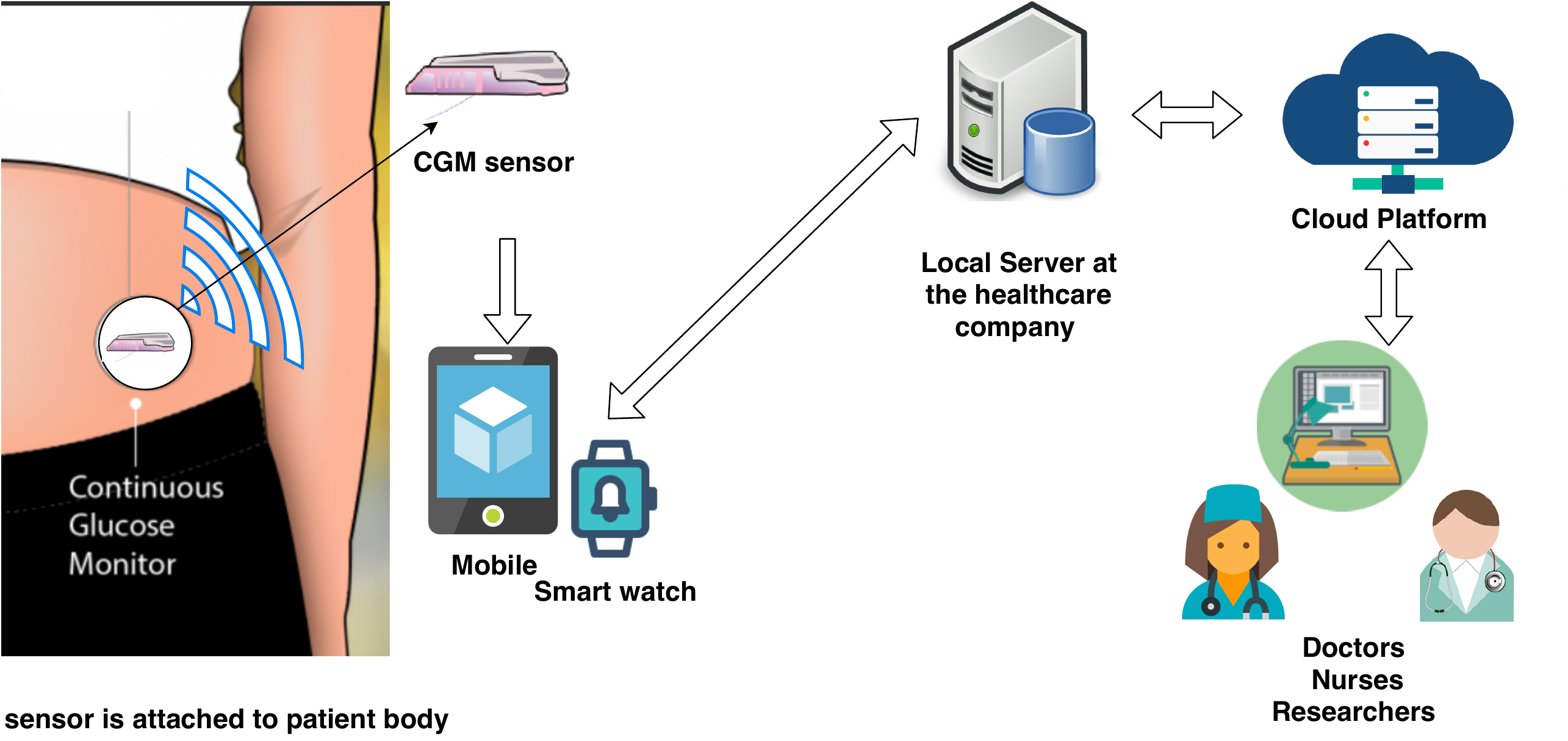}
		\caption{IoT application to support diabetes treatment monitoring}
		\label{Fig:CGM_uscase}
	\end{figure}

	Sara is a researcher in a healthcare company where patients with diabetes require treatment and continual monitoring. Sara is concerned about gathering and analysing data from a Continuous Glucose Monitor (CGM) device worn by patients where the sensor placed into the patient,s body, not into his bloodstream as seen in Figure \ref{Fig:CGM_uscase}. The sensor measures the glucose in the patient interstitial fluid by taking readings at regular intervals for several days. Sara has a monitoring application that can recognize any triggers or patterns for glucose abnormal levels. This application can analyse patient data and produce an alarm to notify the patient and the nurse. There is a speciality nurse that has a level of access to patient data for following-up and provide essential instructions when required. These instructions may include suitable insulin dosage, an exercise plan, daily meals\textbackslash snacks, and types\textbackslash dosage of medications.\\

	\section{IoT Software Development life cycle}\label{Section:IotDLC}
	
	IoT has its own properties, requirements and challenges. Heterogeneous is one of the main challenges where managing and securing different objects, devices, sensors, protocols and applications are complicated \cite{GonzalezGarcia2014}. Moreover, each object could be developed in a diverse way, by diverse manufacturers. As a result, adding some changes or enhancements to the traditional SDLC are required. Privacy as one of the NFRs could be the greatest trouble in a ubiquitous computing \cite{Hong2017}.  Pew Internet research \cite{JANLAURENBOYLES2012} stated that 54 \% of app users refuse to install a mobile app when they realized the amount of personal information the app gathered. Also, 30 \% of the smartphone owners do not want to share personal information which causes uninstalling the app after knowing that it collects personal data.
	
	\begin{figure}
		\includegraphics[scale=0.45]{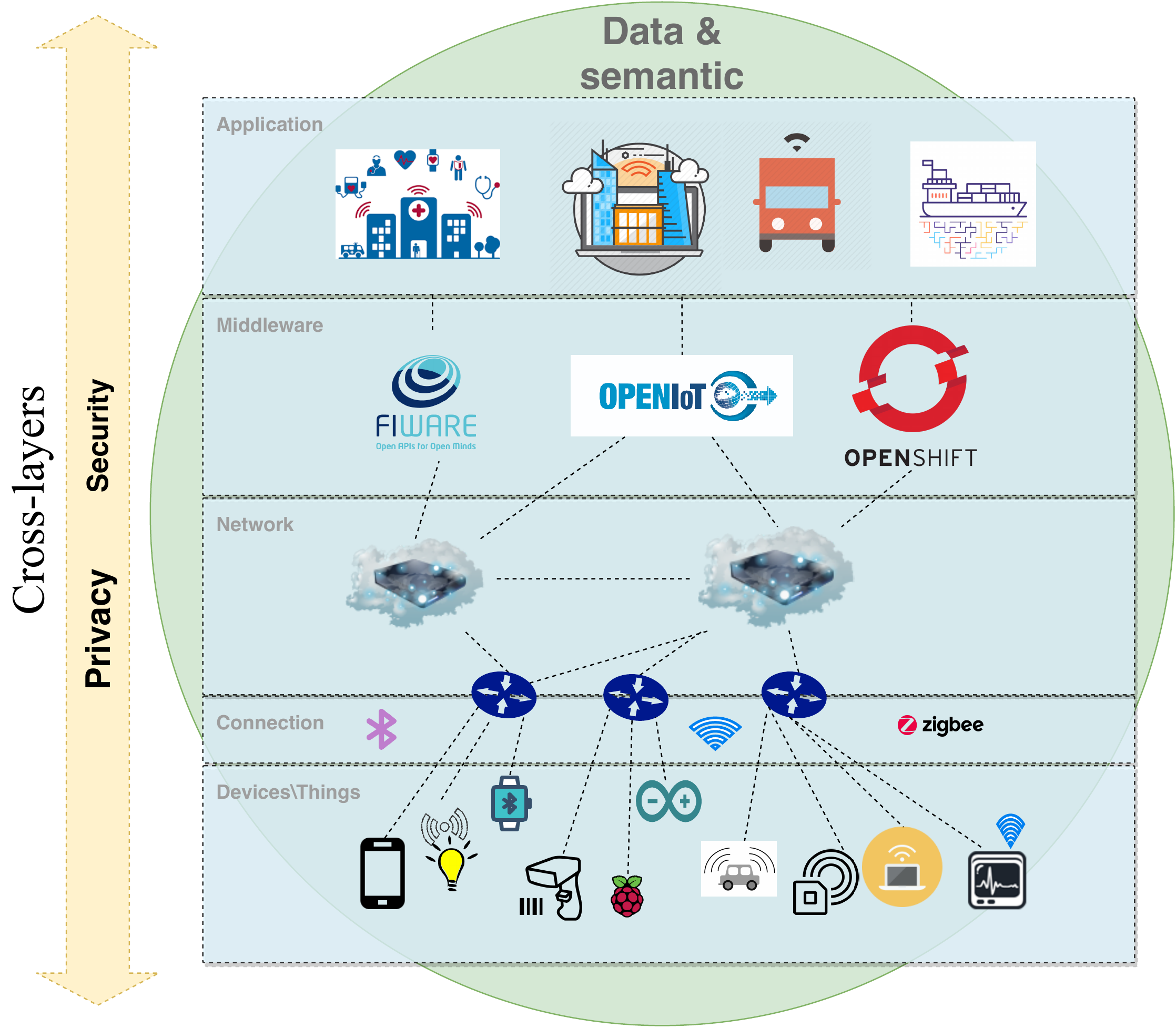}
		\caption{Heterogeneous IoT Framework. Data and semantic  travelling is not straight forward. Starting from the source, it goes through different nodes until it reaches its final destination.}
		\label{Fig:IotFramework}
		\vspace*{-5mm}
	\end{figure}

	We will apply some of the privacy as one of NFRs to the previous use case mentioned in section \ref{sec:CGM_uscase} Figure \ref{Fig:CGM_uscase} to see how these requirements are important in IoT. When the sensor of Continuous Glucose Monitor (CGM) patient's device sends the glucose readings to the application, the data travelling is not straight forward. These data will go through different nodes until it reaches its final destination as seen in Figure \ref{Fig:IotFramework}. One of these nodes could be a third party who could have access to patient sensitive data such as location without having the patient permission or without the research company knowledge. This can happen as a result of the app developers lack of knowing what third parties libraries are collecting from the users \cite{balebako2014privacy} \cite{agarwal2013protectmyprivacy}. Data subjects actually lose control over their data when they are stored on a server operated by a third party. According to \cite{Hong2017}, half of the analyzed app by the study team via PrivacyGrade.org are using location, not because the app wants it but because the third party library uses it. 
	
	In the case of the health field, such as the described use case, distributing patients, data such as location, phone number, patient file number and medical history is critical.
	In this use case anonymization as one of the privacy characteristics is violated. These problems can be solved by applying some of the privacy patterns during software development such as using \textit{protection against tracking, onion routing, and anonymity set} privacy patterns.
	
	Since smartphones are the most developed app in ubiquitous computing, one of the offered solutions is dividing the smartphone ecosystem for privacy into many entities \cite{Hong2017}. These entities share the responsibility for privacy which are:  Developers, Third-party developers, Service providers, App stores, OS providers, Hardware manufacturers, Government agencies and third parties interested in privacy, and Users. Therefore, it very clear IoT applications have a significant potential (much more than mobile applications) to collect personal information that could lead to significant privacy violation. Therefore, it is extremely important to integrate the non-functional requirement into IoT application designs.

	\section{Functional and non functional requirements}\label{Section:FRs_NFRS}
	
	Let us now briefly introduce non-functional requirements in general, before we focus on privacy and security in subsequent sections. Some studies have shown that the effectiveness of requirements engineering (RE), in software projects, is an important success factor \cite{Nasir20112174}. There are many types of requirements and the three common types are: 
	business requirements, the software value in business terms; functional requirements (FRs), known as qualitative requirements; and non-functional requirements (NFRs), known as quantitative requirements \cite{REDDY2011105} \cite{MAXIM201619}.
	Business requirements which define how the system improve the needs of the organization. FRs define the system services, functions or behaviour, that the system supposed to accomplish. On the other hand, NFRs can be defined as the quality attributes (e.g., security, integrity, reliability, usability) or the applied constraints on the application during the development process \cite{Gomariz-Castillo2018} \cite{TerBeek2019}. Most of the efforts that have been done previously are focused on functional requirements (FRs). However, many efforts are focusing currently on non-functional requirements (NFRs) since many studies show that engaging NFRs in early design phases has improved significantly the end-user satisfaction \cite{5636894} \cite{Gomariz-Castillo2018}.

	Functional requirements (FRs) defines the way to understand what is required to build a system correctly and deliver what the end-user expects. In API development, FRs define the expected functionality for the API which requires the end-user collaboration. Functional requirements gathering is located at in the analysis phase \cite{REDDY2011105}. Holding interviews, meetings, or using questionnaires to ask end-users are some of the ways to capture FRs.
	These requirements are typically given a unique identifier for each one and a description in a requirements document. An example of these requirements is:\emph{ REQ1.1. The system shall authenticate that the entered PIN number by the user is correct}.

	\begin{table}[H]
		\caption{high-Level Non-Functional Requirements (NFRs) Classification}
		\label{high-LevelNFRS}
		\footnotesize 
		\begin{tabular}{p{2.25cm}|p{5.5cm}|p{3cm}p{0cm}}
			Characteristics                             & Description                                                                       & Sub-Characteristics     &  \\
			\toprule
			
			\cellcolor[HTML]{ECF4FF}Intrinsic Qualities 
			& characteristics of the product/solution itself                                    
			& Functional suitability  &             \\
			\cellcolor[HTML]{ECF4FF}                    &                                                                                    & Performance efficiency  &            \\
			\cellcolor[HTML]{ECF4FF}                    &                                                                                   & Compatibility           &             \\
			\cellcolor[HTML]{ECF4FF}                    &                                                                                   & Usability               &             \\
			\cellcolor[HTML]{ECF4FF}                    &                                                                                   & Reliability             &             \\
			\cellcolor[HTML]{ECF4FF}                    &                                                                                   & Security                &             \\
			\cellcolor[HTML]{ECF4FF}                    &                                                                                   & Maintainability         &             \\
			\cellcolor[HTML]{ECF4FF}                    &                                                                                   & Portability             &             \\\cline{3-3}
			\cellcolor[HTML]{EFEFEF}Usage Qualities     & characteristics related to outcomes of user  & Effectiveness           &             \\
			\cellcolor[HTML]{EFEFEF}                    & interaction with the product/solution                                                                                   & Efficiency              &             \\
			\cellcolor[HTML]{EFEFEF}                    &                                                                                   & Satisfaction            &             \\
			\cellcolor[HTML]{EFEFEF}                    &                                                                                   & Safety                  &             \\
			\cellcolor[HTML]{EFEFEF}                    &                                                                                   & Usability scope         &             \\\cline{3-3}
			\cellcolor[HTML]{ECF4FF}External Qualities  & market-related characteristics associated               & Service Cost            &             \\
			\cellcolor[HTML]{ECF4FF}                    & with the product/solution                                                                                   & Vendor Risk Mitigation  &             \\
			\cellcolor[HTML]{ECF4FF}                    &                                                                                   & Product Risk Mitigation &             \\
			
			\bottomrule       
		\end{tabular}
	\end{table}

	
	Non-functional requirements are used to support the functional one.
	While FRs describe how the system should behave, the NFRs describe the functioning constraints that guarantee end-user satisfaction. Examples of non-functional requirements include many characteristics such as: performance, flexibility, platform compatibility, security, scalability, usability and recovery \cite{REDDY2011105}.
	ISO/IEC (the International Organization for Standardization) and (the International Electrotechnical Commission) proposes solution attributes, called 'Qualities`, that can be divided into 3 categories in ISO/IEC 25010:2011 \cite{SO/IECJTC1/SC}. These Qualities serve as top-level non-functional requirements (NFRs), which are breakdown to detailed level NFRs as seen in Table \ref{high-LevelNFRS} and \ref{detailed-LevelNFRS}. Privacy framework introduced in another standard called ISO/IEC 29100:2011(en) \cite{272011}. There are many standards from ISO/IEC that are divided based on the policy maker viewpoint such as 29151: Code of practice for personally identifiable information protection and 20889:Privacy enhancing data de-identification techniques. There are many standardization organization and not limited to ISO for defining NFRs, but the listed one here just to have an idea about some them and their meaning from ISO.

	{\footnotesize 
		\begin{longtable}[H]{p{2.25cm}|p{2.75cm}|p{7.75cm}}

			Characteristics                                                  & Sub-Characteristics             & Description \\
			\toprule
			\cellcolor[HTML]{ECF4FF}                                         & Functional completeness         &      degree to which the set of functions covers all the specified tasks and user objectives       \\
			\cellcolor[HTML]{ECF4FF}                                         & Functional correctness          & degree to which a product or system provides the correct results with the needed degree of precision.
			\\
			\multirow{-3}{*}{\cellcolor[HTML]{ECF4FF}Functional suitability} & Functional appropriateness      &         degree to which the functions facilitate the accomplishment of specified tasks and objectives.
			\\
			\cline{3-3}

			\cellcolor[HTML]{EFEFEF}                                         & Time behaviour                  &      degree to which the response and processing times and throughput rates of a product or system, when performing its functions, meet requirements.       \\
			\cellcolor[HTML]{EFEFEF}                                         & Resource utilization            &   degree to which the amounts and types of resources used by a product or system, when performing its functions, meet requirements          \\
			\multirow{-3}{*}{\cellcolor[HTML]{EFEFEF} {\scriptsize Performance efficiency}} & Capacity                        &       degree to which the maximum limits of a product or system parameter meet requirements.
			\\
			\cline{3-3} 
			
			\cellcolor[HTML]{ECF4FF}                                         & Co-existence                    &      degree to which a product can perform its required functions efficiently while sharing a common environment and resources with other products, without detrimental impact on any other product.
			\\
			\multirow{-2}{*}{\cellcolor[HTML]{ECF4FF}Compatibility}          & Interoperability                &    degree to which two or more systems, products or components can exchange information and use the information that has been exchanged.
			\\
			
			\cline{3-3}

			\cellcolor[HTML]{EFEFEF}                                         & Appropriateness recognizability &          degree to which users can recognize whether a product or system is appropriate for their needs.
			\\
			\cellcolor[HTML]{EFEFEF}                                         & Learnability                    & degree to which a product or system can be used by specified users to achieve specified goals of learning to use the product or system with effectiveness, efficiency, freedom from risk and satisfaction in a specified context of use.           \\
			\cellcolor[HTML]{EFEFEF}                                         & Operability                     &   degree to which a product or system has attributes that make it easy to operate and control.
			\\
			\cellcolor[HTML]{EFEFEF}                                         & User error protection           &      degree to which a system protects users against making errors.
			\\
			\cellcolor[HTML]{EFEFEF}                                         & User interface aesthetics       &   degree to which a user interface enables pleasing and satisfying interaction for the user.
			\\
			\multirow{-6}{*}{\cellcolor[HTML]{EFEFEF}Usability}              & Accessibility                   &   degree to which a product or system can be used by people with the widest range of characteristics and capabilities to achieve a specified goal in a specified context of use.
			\\
			\cline{3-3} 
			
			\cellcolor[HTML]{ECF4FF}                                         & Maturity                        &     degree to which a system, product or component meets needs for reliability under normal operation.
			\\
			\cellcolor[HTML]{ECF4FF}                                         & Availability                    &     degree to which a system, product or component is operational and accessible when required for use.
			\\
			\cellcolor[HTML]{ECF4FF}                                         & Fault tolerance                 &       degree to which a system, product or component operates as intended despite the presence of hardware or software faults.
			\\
			\multirow{-4}{*}{\cellcolor[HTML]{ECF4FF}Reliability}            & Recoverability                  &     degree to which, in the event of an interruption or a failure, a product or system can recover the data directly affected and re-establish the desired state of the system.
			\\
			\cline{3-3}

			\cellcolor[HTML]{EFEFEF}                                         & Confidentiality                 &   degree to which a product or system ensures that data are accessible only to those authorized to have access.
			\\
			\cellcolor[HTML]{EFEFEF}                                         & Integrity                       &      degree to which a system, product or component prevents unauthorized access to, or modification of, computer programs or data.
			\\
			\cellcolor[HTML]{EFEFEF}                                         & Non-repudiation                 &         degree to which actions or events can be proven to have taken place, so that the events or actions cannot be repudiated later.
			\\
			\cellcolor[HTML]{EFEFEF}                                         & Accountability                  &     degree to which the actions of an entity can be traced uniquely to the entity.
			\\
			\multirow{-5}{*}{\cellcolor[HTML]{EFEFEF}Security}               & Authenticity                    &     degree to which the identity of a subject or resource can be proved to be the one claimed.
			\\
			\cline{3-3}

			\cellcolor[HTML]{ECF4FF}                                         & Modularity                      &     degree to which a system or computer program is composed of discrete components such that a change to one component has minimal impact on other components.
			\\
			\cellcolor[HTML]{ECF4FF}                                         & Reusability                     &    degree to which an asset can be used in more than one system, or in building other assets.
			\\
			\cellcolor[HTML]{ECF4FF}                                         & Analysability                   &        degree of effectiveness and efficiency with which it is possible to assess the impact on a product or system of an intended change to one or more of its parts, or to diagnose a product for deficiencies or causes of failures, or to identify parts to be modified.
			\\
			\cellcolor[HTML]{ECF4FF}                                         & Modifiability                   &        degree to which a product or system can be effectively and efficiently modified without introducing defects or degrading existing product quality.
			\\
			\multirow{-5}{*}{\cellcolor[HTML]{ECF4FF}Maintainability}        & Testability                     &   degree of effectiveness and efficiency with which test criteria can be established for a system, product or component and tests can be performed to determine whether those criteria have been met.
			\\
			\cline{3-3}

			\cellcolor[HTML]{EFEFEF}                                         & Adaptability                    &        degree to which a product or system can effectively and efficiently be adapted for different or evolving hardware, software or other operational or usage environments.
			\\
			\cellcolor[HTML]{EFEFEF}                                         & Installability                  &     degree of effectiveness and efficiency with which a product or system can be successfully installed and/or uninstalled in a specified environment.
			\\
			\multirow{-3}{*}{\cellcolor[HTML]{EFEFEF}Portability}            & Replaceability                  &        degree to which a product can replace another specified software product for the same purpose in the same environment.
			\\
			
			\cline{3-3}

			\cellcolor[HTML]{ECF4FF}                                         &  anonymization                   &   process by which personally identifiable information (PII) is irreversibly altered in such a way that a PII principal can no longer be identified directly or indirectly, either by the PII controller alone or in collaboration with any other party
			\\
			\cellcolor[HTML]{ECF4FF}                                         &           pseudonymization        &     process applied to personally identifiable information (PII) which replaces identifying information with an alias
			\\
			\multirow{-3}{*}{\cellcolor[HTML]{ECF4FF}Privacy *}            &        consent           &        personally identifiable information (PII) principal,s freely given, specific and informed agreement to the processing of their PII
			\\
			\bottomrule

			\caption{Detailed-Level of Non-Functional Requirements (NFRs) Classification from ISO/IEC 25010 \cite{SO/IECJTC1/SC}. (*)Privacy characteristics are example and are not limited to the listed one here from ISO/IEC 29100 \cite{272011}.}
			\label{detailed-LevelNFRS}  
			
		\end{longtable}
		
	}

	
	\section{Methodology}\label{Section:Methodology}
	In order to build this survey, we followed a search strategy. The first step was selecting papers in google scholar to avoid publisher basis. Then we search on specific libraries based on forward and backword snowballing. After that, it was data extraction step to extract some of the properties from each notation/representation. Finally, we analysed the data to find the review results.
	
	\begin{figure}
		\includegraphics[scale=0.8]{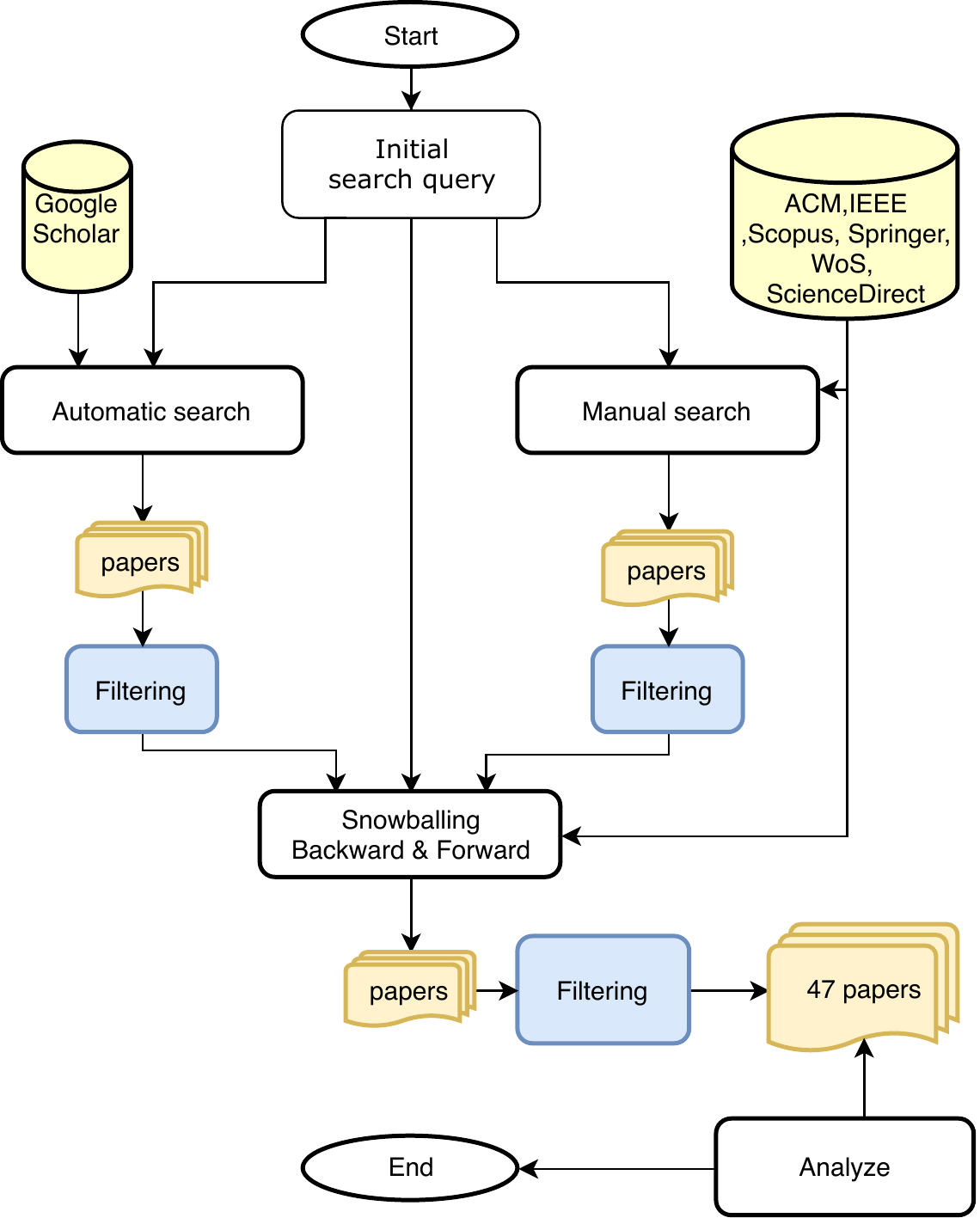}
		\caption{Search Process}
		\label{searchProcess}
	\end{figure}
	
	\begin{figure}
		\includegraphics[scale=0.65]{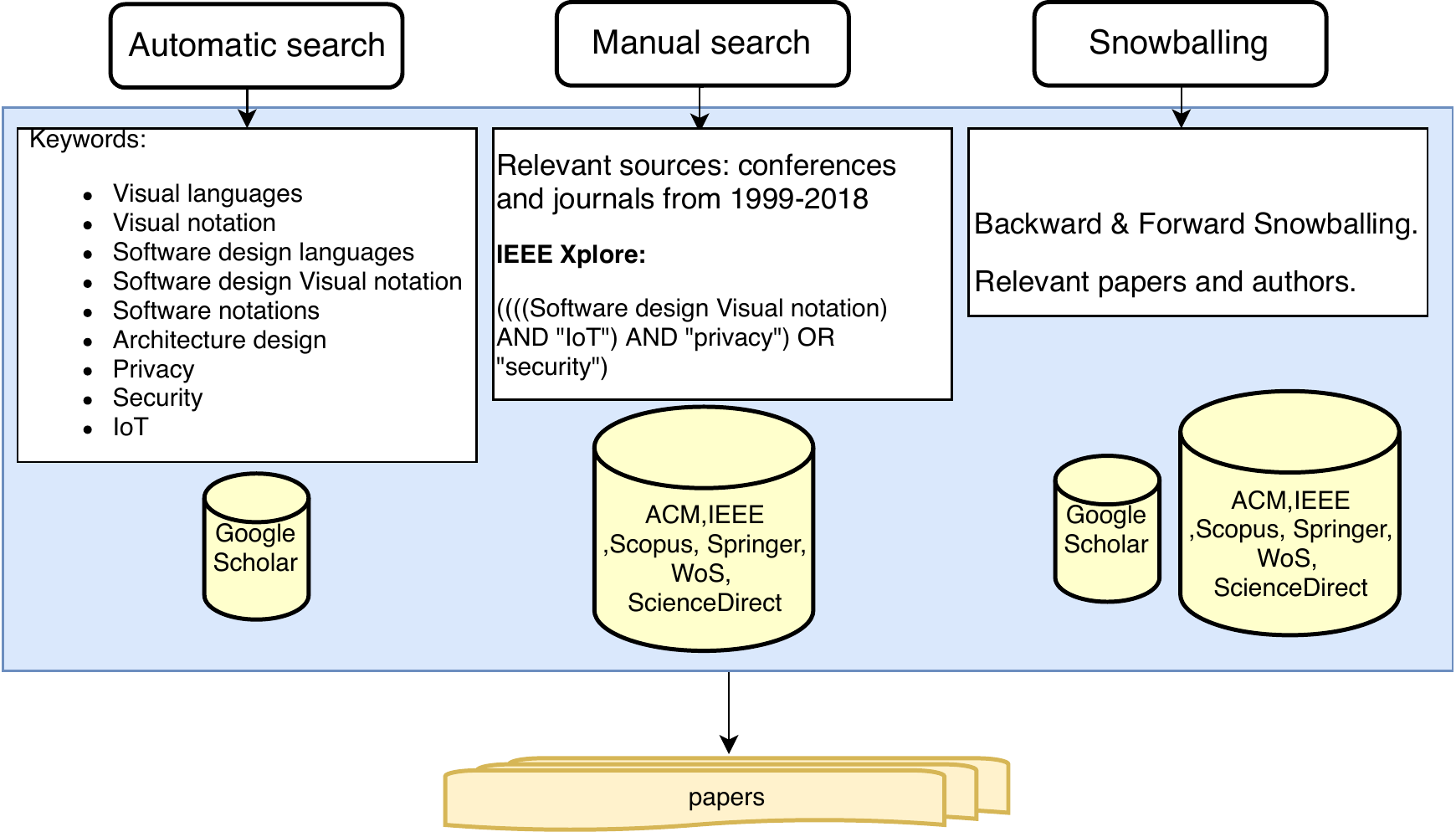}
		\caption{Initial search}
		\label{InitialsearchProcess}
	\end{figure}
	
	\subsection{Data Sources and Search Strategy}
	For data collection and extraction, Kitchenham method \cite{Kitchenham2013} is generally (not all the steps) used as a guideline to extract data from each paper.
	At the beginning, we create the initial search query by using Google Scholar, using some keywords that include the word  notation or non-functional requirements (security and privacy in specific). This resulted in general queries with combination of AND and OR in between as seen in Table \ref{Queries}. These queries resulted in having many papers where some of them is not very related to visual notation or non-functional requirements. However, this step is used to give us an idea about which digital libraries and journals that interested in notation, visualization and non-functional requirements. After that, we tried hybrid search using more complex quires in specific journals and libraries which listed in Table \ref{Libraries} such as IEEE Xplore , Scopus, Springer, ACM, ScienceDirect etc. 
	
	For the notations study, papers from 1999 onward have been checked to ensure comparability and to cover wide varieties of notations in general and secure one in specific. The search process is done in many stages and several iterations that include three search processes: automatic, manual, and snowballing.
	
	\begin{itemize}
		
		\item Automatic search (Figure \ref{searchProcess} and \ref{InitialsearchProcess}): This stage was performed using search engine using keywords that combined any of the terms \textit{`Visual languages', `Visual notation', `Software design languages', `Software design Visual notation', `Software notations', Architecture design, `Privacy', `use all above combinations', `Security', `IoT', `Architecture', `cyber physical systems' and `ubiquitous Computing'}. Google Scholar is used as a start point for searching since it is considered as a good substitute to avoid bias in favour of any specific publisher \cite{wohlin2014guidelines}. Using this way, we took the advantage looking for the whole spectrum for all the available publications regardless of the publishers. 
		
		\item Manual search (Figure \ref{searchProcess} and \ref{InitialsearchProcess}): This stage was done using the proceedings of some conferences and journals as sources such as Ubiquitous Computing (UbiComp), Journal of Systems and Software (JSS), Transactions on Software Engineering and Methodology (TOSEM) and others listed in Table \ref{Libraries}. For these sources, the studied time period is 2000-2019.

		\item Snowballing (Figure \ref{searchProcess} and \ref{InitialsearchProcess}): This stage was performed on the set of papers that gathered previously in manual search and based on known papers from the same relevant authors and time period. Then, backward snowballing is performed by checking the references to select relevant papers based on relevant title, abstract, and general structure review.
	\end{itemize}

	\begin{table}[]
		\centering
		\scriptsize
		\caption{Some queries and terms for online libraries search.}
		\label{Queries}	
		\begin{tabular}{l|l}
			Category                             & Queries and Terms                                                                                                                                                                                                            \\\toprule
			\cellcolor[HTML]{DAE8FC}General      & ``Architecture design''                                                                                                                                                                                     \\
			\cellcolor[HTML]{DAE8FC}             &``Visual'' AND (`` languages'' OR ``notation'')                                                                                                                                                                                                        \\
			\cellcolor[HTML]{DAE8FC}             & ``Software design languages'' OR ``Software design Visual notation''                                                                                                                                                             \\
			\cellcolor[HTML]{DAE8FC}             & ``Security'' AND (combinations of the above)                                                                                                                                                                                   \\
			\cellcolor[HTML]{DAE8FC}             & ``Privacy'' AND (combinations of the above)                                                                                                                                                                                                                                                                                                                                 \\
			
			\cellcolor[HTML]{FFFFC7}&\\
			\cellcolor[HTML]{FFFFC7}More spastic             & \textbf{IEEE:} ((((Software design Visual notation) AND ``IoT'') AND ``privacy'') OR ``security'')\\

			\cellcolor[HTML]{FFFFC7}             &	\textbf{ACM:} { (+Visual +notation software requirements +security +privacy)} \\
			
			\cellcolor[HTML]{FFFFC7} & \textbf{Scopus:} TITLE-ABS-KEY-AUTH(``non functional'' AND requirements AND IoT) \\
			\cellcolor[HTML]{FFFFC7}  & AND ( LIMIT-TO ( SUBJAREA,``COMP'' )
			OR LIMIT-TO ( SUBJAREA,``ENGI'' ) )\\ 
			\cellcolor[HTML]{FFFFC7}  & AND ( LIMIT-TO ( LANGUAGE,``English'' ) ) \\\bottomrule                                                                                                                      \end{tabular}
	\end{table}

	\begin{table}[]
		\centering
		\tiny
		\caption{Sources of Selected conference proceedings and journals for Manual and Automatic search that the notations are selected from.}
		\label{Libraries}
		\begin{tabular}{@{\extracolsep{\fill}}p{1.5cm}p{1.25cm} p{7.5cm}p{1.5cm}}	
			\toprule
			{\scriptsize Venue }& {\scriptsize Abbr. }              &  {\scriptsize Source}  &{\scriptsize Publisher}  \\
			\bottomrule
			\cellcolor[HTML]{ECF4FF}                              & OOPSLA               & ACM SIGPLAN conference companion on Object Oriented Programming Systems Languages and Applications                   & ACM                     \\
			\cellcolor[HTML]{ECF4FF}                              & ESORICS              & European Symposium on Research in Computer Security                                                                  & Springer                \\
			\cellcolor[HTML]{ECF4FF}                              & DAC                  & IEEE Design Automation Conference                                                                                    & ACM\textbackslash{}IEEE \\
			\cellcolor[HTML]{ECF4FF}                              & EuroS PW             & IEEE European Symposium on Security and PrTransactions on Software Engineeringivacy Workshops                                                            & IEEE                    \\
			\cellcolor[HTML]{ECF4FF}                              & COMPSAC              & IEEE International Computer Software and Applications Conference                                                     & IEEE                    \\
			\cellcolor[HTML]{ECF4FF}                              & ICECCS               & IEEE International Conference on Engineering Complex Computer Systems Navigating Complexity in the e-Engineering Age & IEEE                    \\
			\cellcolor[HTML]{ECF4FF}                              & ICECCS               & IEEE International Conference on Engineering of Complex Computer Systems                                             & IEEE                    \\
			\cellcolor[HTML]{ECF4FF}                              & SCC                  & IEEE International Conference on Services Computing                                                                  & IEEE                    \\
			\cellcolor[HTML]{ECF4FF}                              & ICSA-C               & IEEE International Conference on Software Architecture Companion                                                     & IEEE                    \\
			\cellcolor[HTML]{ECF4FF}                              & ICWS                 & IEEE International Conference on Web Services                                                                        & IEEE                    \\
			\cellcolor[HTML]{ECF4FF}                              & ISESS                & IEEE International Software Engineering Standards Symposium and Forum                                                & IEEE                    \\
			\cellcolor[HTML]{ECF4FF}                              & VL/HCC               & IEEE Symposium on Visual Languages-Human Centric Computing                                                           & IEEE                    \\
			\cellcolor[HTML]{ECF4FF}                              & AVI                  & International Conference on Advanced Visual Interfaces                                                               & ACM                     \\
			\cellcolor[HTML]{ECF4FF}                              & ICSE                 & International Conference on Software Engineering                                                                     & ACM\textbackslash{}IEEE \\
			\cellcolor[HTML]{ECF4FF}                              & SEKE                 & International Conference on Software Engineering and Knowledge Engineering                                           & Springer                \\
			\multirow{-16}{*}{\cellcolor[HTML]{ECF4FF}Conference} & FME                  & International FME Workshop on Formal Methods in Software Engineering                                                 & ACM\textbackslash{}IEEE \\
			\cellcolor[HTML]{EFEFEF}                              & IST                  & Information and Software Technology                                                                                  & Elsevier                \\
			\cellcolor[HTML]{EFEFEF}                              & -                     & Computers \& Security                                                                                                & Elsevier                \\
			\cellcolor[HTML]{EFEFEF}                              &-                      & Computer Networks                                                                                                    & Elsevier                \\
			\cellcolor[HTML]{EFEFEF}                              &-                      & Decision Support Systems                                                                                             & Elsevier                \\
			\cellcolor[HTML]{EFEFEF}                              & TSE                  & IEEE Transactions on Software Engineering                                                                      & IEEE                    \\
			\cellcolor[HTML]{EFEFEF}                              & IEEE T SYST MAN CY C & IEEE Transactions on Systems, Man, and Cybernetics, Part C (Applications and Reviews)                                & IEEE                    \\
			\cellcolor[HTML]{EFEFEF}                              & IST                  & Information and Software Technology                                                                                  & Elsevier                \\
			\cellcolor[HTML]{EFEFEF}                              & IJSE                 & International Journal of Software Engineering                                                                        & CSC                     \\
			\cellcolor[HTML]{EFEFEF}                              & INTR                 & Internet Research                                                                                                    & Emerald                 \\
			\cellcolor[HTML]{EFEFEF}                              & JSW                  & Journal of Software                                                                                                 & -                       \\
			\cellcolor[HTML]{EFEFEF}                              & -                    & Journal of Visual Languages and Computing                                                                            & Elsevier                \\
			\cellcolor[HTML]{EFEFEF}                              & -                    & Science of Computer Programming                                                                                      & Elsevier                \\
			\cellcolor[HTML]{EFEFEF}                              & SoSyM                & Software and Systems Modeling                                                                                        & Springer                \\
			\multirow{-14}{*}{\cellcolor[HTML]{EFEFEF}journal}    & -                    & The Journal of Systems and Software                                                                                  & Elsevier \\
			\bottomrule              
		\end{tabular}
	\end{table}

	
	\subsection{Data Extraction}
	
	As stated previously, Kitchenham method is generally followed for data clollecting. For each paper many characteristics were extracted. All the collected and extracted data was structured in an Excel spread sheet. For this study, a repository has been built which contains the meta-data for the analyzed papers. Based on the built repository, we accomplished our analysis of the analyzed papers. The meta-data contains: notation ID,  name,  publication year, scope,  visual design based,  tool support,  security/privacy support, IoT support, validation type, experiments' participants background,  and the conference proceeding/journal. Finally, a list of 47 notations in different publications remained, and all of them were analyzed in this survey.

	%
	%

	\subsubsection*{Analysis of the apers}
\begin{figure}
		\includegraphics[scale=0.5]{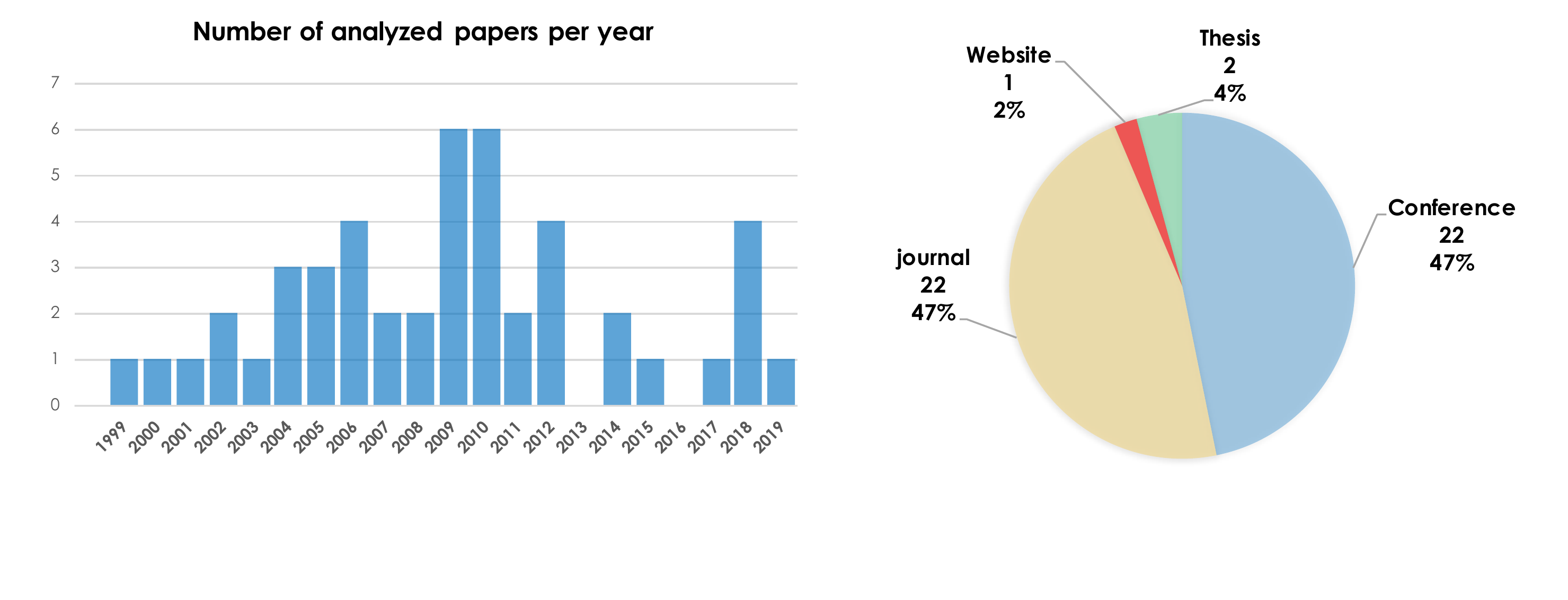}
		\caption{Overview of the number of analysed  papers. (a)  distributed per year (1999-2019), and (b) by venue.}
		\label{Overview_notation_venue}
		\vspace*{-2.5mm}
	\end{figure}

	The publications distribution per year from 1999 to 2019 is presented in Figure \ref{Overview_notation_venue} (A).  The distribution of all analyzed papers' publication types can be found in Figure \ref{Overview_notation_venue} (B). It can be noticeable that the peak publications on secure notation was in 2009-2010 with 12 papers in total. In addition, most of the found papers come from journals and conference proceedings.

	\section{Design Notation, Languages, and Representations}\label{Section:RESULTS_OF_THE_REVIEW}

	Many design notations have been created for security that intended to document security concepts and features into a software design model. These efforts about notations are distributed and not organized. This distribution makes it complex for researchers to assist existing notations to decide which technique they need to follow. There is an effort has been done in the field of reviewing security notations such as \cite{VandenBerghe2017}; however, it does not have anything about IoT and it needs to be updated.

	The following section presents a systematic literature review that investigates the available non-functional requirements (security and privacy) notations and produces a comprehensive analysis for each one of them. This will be done to understand which the techniques are they used to represent security characteristics visually. After that, we will observe which privacy technique we can match to develop later a model for a privacy notation.
	
\begin{figure}
		\includegraphics[scale=0.35]{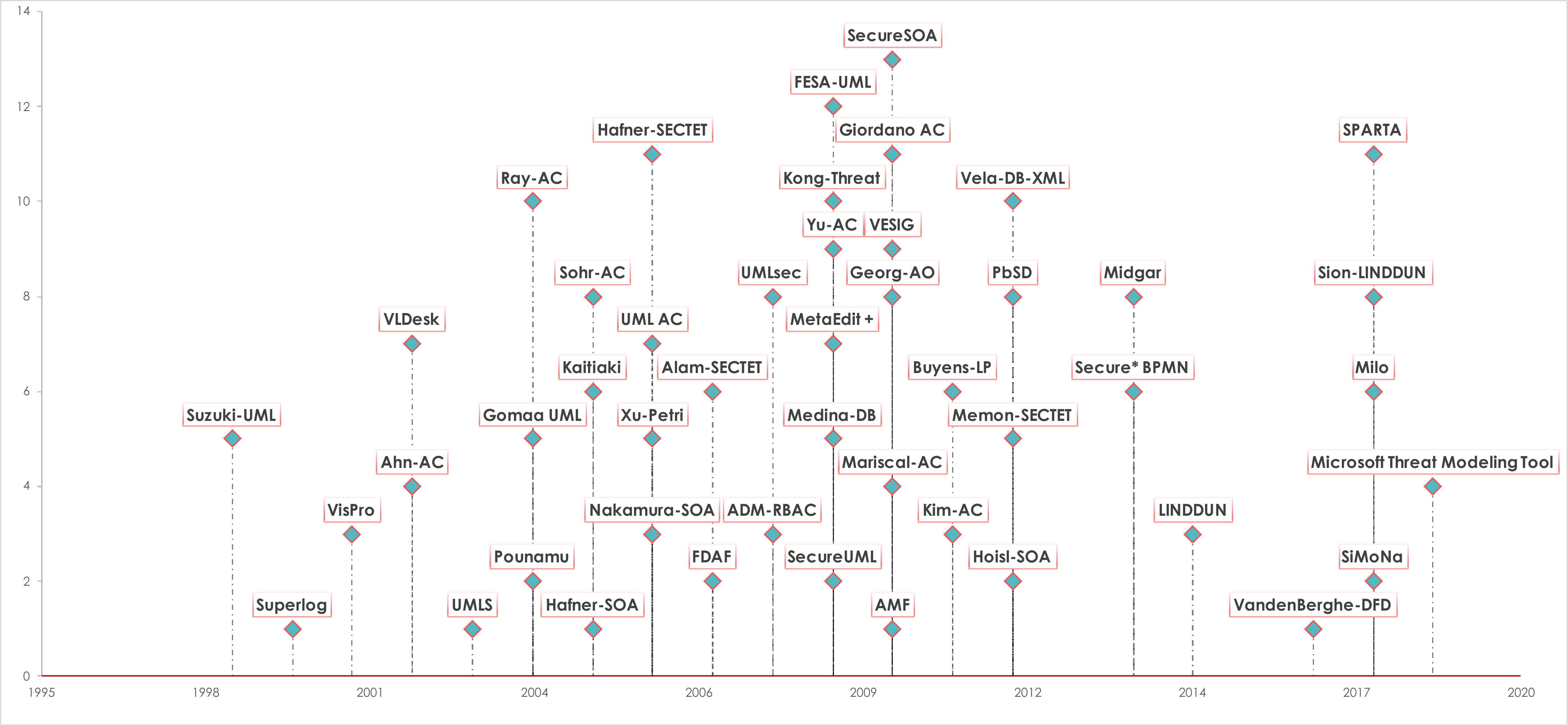}
		\caption{Timeline for the analyzied notations, models, languages from 1999 to 2019.}
		\label{fig:timeline99-18}
	\end{figure}

\begin{figure}[p]
		\scriptsize
		
		\centering
		
		\vspace{-13pt}
		\makebox[\linewidth]{
			\includegraphics[angle=90, scale=.3, keepaspectratio]{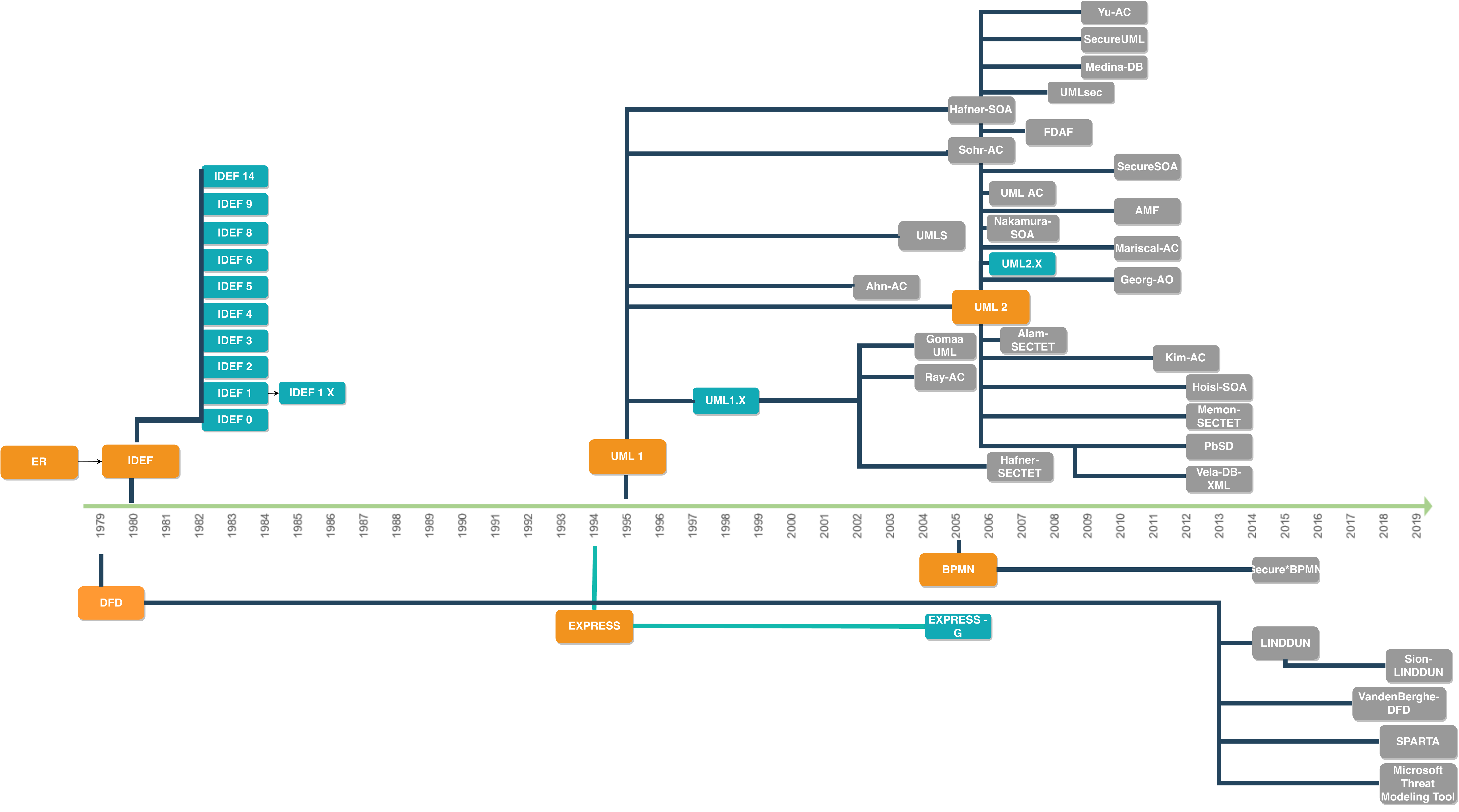}
			
		}
		\caption{The analyzied notations, models, languages over the time ( in grey rectangles) with their associated standard notations such as UML, DFD and BPMN (orange rectangles).}
		\label{fig:Timeline}
	\end{figure}

	In the beginning, we analyze 47 notations, languages, and representations for the period (1999-2019) to have an overview of design representations over the last 20 years as seen in Figure \ref{fig:timeline99-18}. These notations could follow some of the known representation standards such as UML and DFD, which most of them do, as seen in Figure \ref{fig:Timeline}. In addition to the representation model, all the notations are investigated in term of scope, coverage and tool support. Then we assessed the way these notations have been evaluated and who are the participates in case the notations are experimentally validated. \\

	\begin{table}[!t]
		\caption{ Analyses of 47 design notations based on the covered scope of notation.}
		\label{scope}
		\tiny	
		\rowcolors{2}{white}{gray!20}
		
		\begin{tabularx}{4.5in}{ ll l|| ccccccccc |ccc}
			\toprule
			\multicolumn{15}{c}{\textbf{List of notations}} \\
			\midrule
			
			\rowcolor{white}
			
			&&&\multicolumn{9}{c|}{\textbf{Scope}}
			&\multicolumn{3}{c}{\textbf{Coverage}}\\
			
			\textbf{Name}
			&\textbf{Citation} 
			& \textbf{Year}

			&\rotatebox{90}{Generic}
			&\rotatebox{90}{Database}
			&\rotatebox{90}{Web system}
			&\rotatebox{90}{SOA}
			&\rotatebox{90}{Authorization}
			&\rotatebox{90}{Patterns design}
			&\rotatebox{90}{Education}
			&\rotatebox{90}{DSML}
			&\rotatebox{90}{System-on-Ship}
			&\rotatebox{90}{IoT}
			&\rotatebox{90}{Security}
			&\rotatebox{90}{Privacy}\\
			
			\bottomrule 
			\toprule
			
			ADM-RBAC & \cite{Diaz2008a} 	&	2008	 		&		&		&\CheckmarkBold 		&		&		&	&&		&&&\CheckmarkBold&\\
			
			Ahn-AC &\cite{Ahn2002}	& 2002	 		&\CheckmarkBold 		&		&		&		&		&	&&		&&&\CheckmarkBold&\\
			
			Alam-SECTET &	\cite{Ahn2002}&	2007	 	&		&		&		&\CheckmarkBold 		&		&	&&		&&&\CheckmarkBold&\\
			
			AMF &\cite{Hu2010}	&	2010	 			&		&		&		&		&\CheckmarkBold 		&	&&		&&&\CheckmarkBold&\\
			
			Buyens-LP &	\cite{Buyens2013} &	2011			&\CheckmarkBold 		&		&		&		&		&	&&		&&&\CheckmarkBold&\\
			
			FDAF &	\cite{Dai2007} &	2007	 			&\CheckmarkBold 		&		&		&		&		&	&&		&&&\CheckmarkBold&\\
			
			Georg-AO &\cite{Georg2009}	&	2010	 		&\CheckmarkBold 		&		&		&		&		&	&&		&&&\CheckmarkBold&\\
			
			Giordano AC &\cite{Giordano2010a}	&	2010	 	&\CheckmarkBold 		&		&		&		&		&	&&		&&&\CheckmarkBold&\\
			
			Gomaa UML &	\cite{Shin2004}&	2004	 		&\CheckmarkBold 		&		&		&		&		&	&&		&&&\CheckmarkBold&\\
			
			Hafner-SOA &\cite{Hafner-2005}	&	2005	 	&		&		&		&\CheckmarkBold 		&		&	&&		&&&\CheckmarkBold&\\
			
			Hoisl-SOA &	\cite{Hoisl2014}&	2012 		&		&		&		&\CheckmarkBold 		&		&	&&		&&&\CheckmarkBold&\\
			
			Kim-AC &	\cite{Kim2011}&	2011	 		&\CheckmarkBold 		&		&		&		&		&	&&		&&&\CheckmarkBold&\\
			
			Kong-Threat &\cite{KONG2010} &2009		 	&\CheckmarkBold 		&		&		&		&		&	&&		&&&\CheckmarkBold&\\
			
			Mariscal-AC &\cite{Pavlich-Mariscal2010}	&	2010	 	&\CheckmarkBold 		&		&		&		&		&	&&		&&&\CheckmarkBold&\\
			
			Medina-DB& \cite{Trujillo2009} &2009		 		&		&\CheckmarkBold 		&		&		&		&	&&		&&&\CheckmarkBold&\\
			
			Memon-SECTET &	\cite{Memon2014} &	2012	 	&		&		&		&\CheckmarkBold 		&		&	&	&	&&&\CheckmarkBold&\\
			
			Nakamura-SOA &\cite{Satoh2006}	&	2006	 	&		&		&		&\CheckmarkBold 		&		&	&&		&&&\CheckmarkBold&\\
			
			PbSD &\cite{Abramov2012}	&	2012	 			&		&\CheckmarkBold 		&		&		&		&	&&		&&&\CheckmarkBold&\\
			
			Ray-AC &\cite{Ray2004}	&	2004 		&\CheckmarkBold 		&		&		&		&		&	&	&	&&&\CheckmarkBold&\\
			
			SecureSOA &	\cite{Menzel2010}&	2010	 		&		&		&		&\CheckmarkBold 		&		&	&&		&&&\CheckmarkBold&\\
			
			SecureUML &	\cite{Basin2009}&	2009	 		&\CheckmarkBold 		&		&		&		&		&	&&		&&&\CheckmarkBold&\\
			
			Sohr-AC &\cite{Sohr2005}	&	2005	 		&\CheckmarkBold 		&		&		&		&		&	&&		&&&\CheckmarkBold&\\
			
			UML AC &\cite{Koch2006}	&	2006	 		&\CheckmarkBold 		&		&		&		&		&	&	&	&&&\CheckmarkBold&\\
			
			UMLS &	\cite{Heldal2003}&	2003	 			&\CheckmarkBold 		&		&		&		&		&	&	&	&&&\CheckmarkBold&\\
			
			UMLsec &\cite{Jurjens2008}	&	2008	 		&\CheckmarkBold 		&		&		&		&		&	&	&	&&&\CheckmarkBold&\\
			
			Vela-DB-XML &\cite{Vela2012}	&	2012	 	&		&\CheckmarkBold 		&		&		&		&	&	&	&&&\CheckmarkBold&\\
			
			Xu-Petri &\cite{Xu2006}	&	2006 		&\CheckmarkBold 		&		&		&		&		&	&	&	&&&\CheckmarkBold&\\
			
			Yu-AC &\cite{Yu2009}	&	2009	 			&\CheckmarkBold 		&		&		&		&		&	&	&	&&&\CheckmarkBold&\\
			
			Hafner-SECTET &\cite{Hafner2006b}	&	2006	 	&		&		&		&\CheckmarkBold		&		&	&	&	&&&\CheckmarkBold&\\
			
			SPARTA &\cite{Sion2018}	&	2018	 		&\CheckmarkBold 		&		&		&		&		&	&	&	&&&\CheckmarkBold&\CheckmarkBold\\

			VandenBerghe-DFD & \cite{VanDenBerghe20173}&2017		&\CheckmarkBold		&		&		&		&		&	&		&	&&&&\\
			LINDDUN	 &\cite{Wuyts2015a}&	2015	 		&\CheckmarkBold		&		&		&		&		&	&		&	&&&&\CheckmarkBold\\
			Sion-LINDDUN	 &\cite{Sion2018b}	&	2018	&\CheckmarkBold		&		&		&		&		&	&		&	&&&&\CheckmarkBold\\
			Kaitiaki &	\cite{Liu2005}&	2005	 		&			&		&		&		&		&	&		&{\color{red}\CheckmarkBold}	&&&&\\
			Suzuki-UML &\cite{Suzuki1999}	&	1999	 	&		&		&\CheckmarkBold			&		&		&	&	&		&&&&\\
			VLDesk &\cite{Francese2004}	
			&	2002	 		&		&		&		&		&	&{\color{red}\CheckmarkBold}			&      &		&&&&\\
			VESIG & \cite{diaz2010visual}	&	2010	 			&		&		&		&		&		&{\color{red}\CheckmarkBold}	&	&		&&&&\\
			Superlog &\cite{Flake2000}	&	2000	 		&		&		&		&		&		&	&			&&\CheckmarkBold&&&\\
			VisPro &\cite{Zhang2001}	&	2001	 		&{\color{red}\CheckmarkBold}		&		&		&		&		&	&			&{\color{red}\CheckmarkBold}&&&&\\
			FESA-UML &\cite{Farkas2009}	&	2009	 		&{\color{red}\CheckmarkBold}		&		&		&		&		&	&		&	&&&&\\
			Pounamu &\cite{NianpingZhu2005}	&	2004	 		&{\color{red}\CheckmarkBold}		&		&		&		&		&	&	&{\color{red}\CheckmarkBold}		&&&&\\
			Milo &\cite{Rao2018}	&	2018				&		&		&		&		&		&&\CheckmarkBold		&	&&&&\\
			SiMoNa &\cite{DeMorais2018}	&	2018	 		&		&		&		&		&		&	&	&\CheckmarkBold	&&\CheckmarkBold&&\\
			Midgar &\cite{GonzalezGarcia2014}&	2014	 		&		&		&		&		&		&	&		&\CheckmarkBold	&&\CheckmarkBold&&\\
			MetaEdit+ 5.5 &	\cite{Tolvanen2009} &	2017	 		&		&		&		&		&		&	&		&\CheckmarkBold	&&&&\\
			Microsoft Threat Modeling &	\cite{Microsoft} &	2019	 		&\CheckmarkBold		&		&		&		&		&	&		&&	&\CheckmarkBold&\CheckmarkBold&\\
			Secure*BPMN &	\cite{Cherdantseva2014}&	2014	 		&	&		&		&		&		&	&		&&	&&\CheckmarkBold&\\
			\midrule	
			\textbf{Total} &		 		& &	25 &	3 &	2 &	7 &	1 &	2 &	1 &	6 &	1 &	3&{\color{red}32} &3\\	
			
			\bottomrule

		\end{tabularx}
	\end{table}

	
	\subsection{Scope and Coverage }

	This study analyses 47 design notations for different coverage and purposes (see Table \ref{scope}). This study is trying to look at a variety of notations' domains to assess how they are built, how they support NFRs and how IoT can be built. Since the study are related to IoT, three of these notations are covering IoT systems which are SiMoNa, Midgar and Microsoft Threat Modeling Tool. Since building IoT systems is expanding rapidly for the last decade, it is expected to see the three systems are recently produced in the last five years; 2014 for Midgar, 2018 for SiMoNa and 2019 for Microsoft Threat Modeling Tool. Regarding the Non-functional requirements (NFRS), security and privacy are the two assessed NFRs in this study. Security has been supported for more than the half of the analysed papers (32  notations out of 47), while privacy is only covered in SPARTA, LINDDUN and Sion-LINDDUN ( 3 notations out of 47).

	It is noticeable that more than half of the analysed notations (25 out of 47 ) are generic and are not focused on a specific application domain. There are some domain-specific notations, and service-oriented architecture (SOA) is relatively common (7 notations). The other notations are vary between database (3 notations), DSML (6 notations), web system(2 notations), design pattern recovery (2 notations), authorization (1 notation), education (1 notation) and system-on-Ship (1 notation). 

	\subsection{Tool support}
\begin{figure}
		\includegraphics[scale=.50]{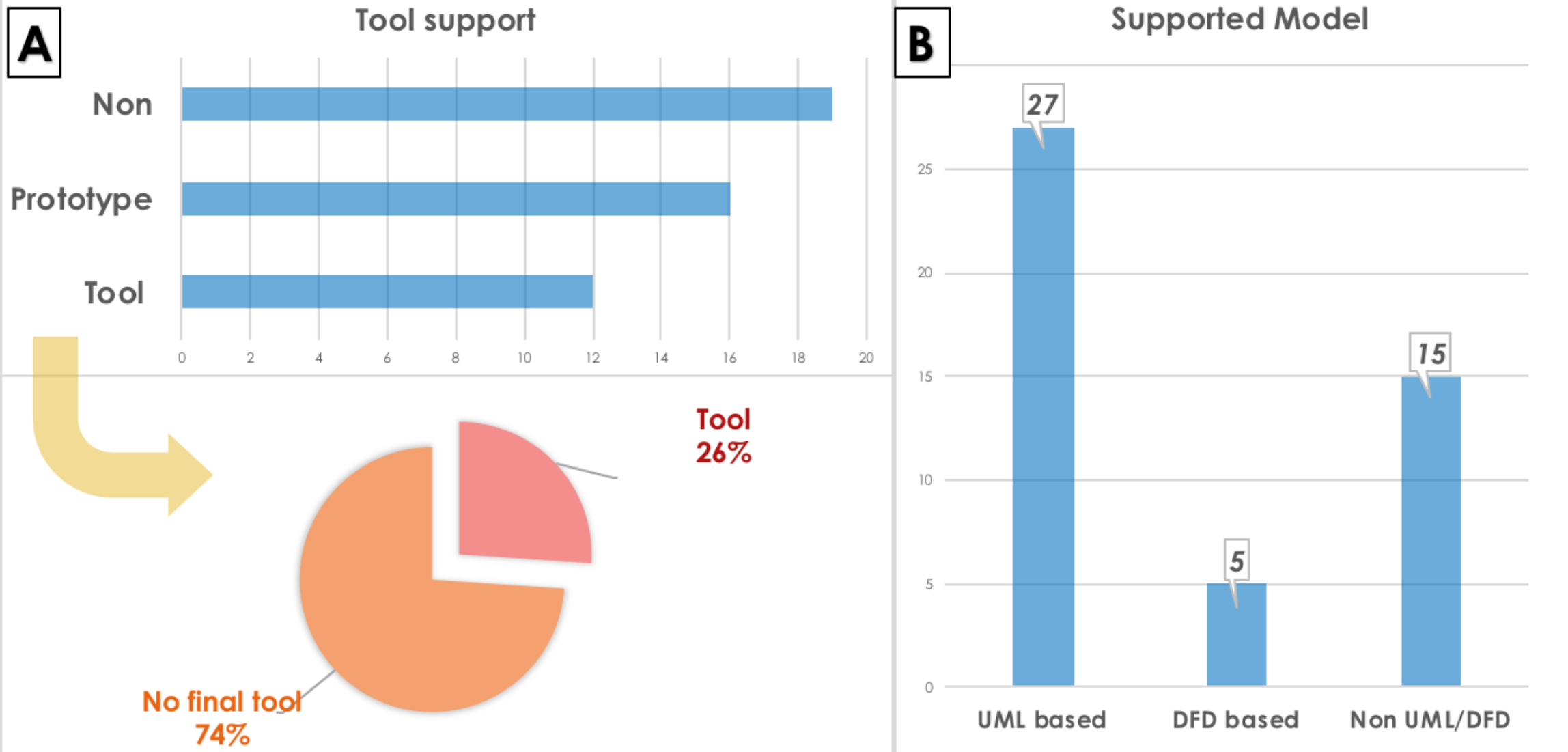}
		\caption{chart (A) shows most of the notations did not produce any tool (13 out of 47). chart (B) shows that most of the analyzed papers are UML-based modelling.}
		\label{tool_modelBased}	
	\end{figure}

	It can be seen from the Table \ref{Poduct_ModelBased} and Figure \ref{tool_modelBased} that the majority of the analyzed notations that they did not produce final tool (34 notations:16 notations with kind of prototype and 18 without any tool support). Only around a quarter of the notations have a tool (12 out of 47 notations) and some of the notations that have a tool do not support security such as SiMoNa which is IoT infographics domain-specific modelling language. As stated in \cite{VandenBerghe2017}, lacking tool support is due to the low maturity of the secure software design field and there is a gap in this area.

	\subsection{Representation support }	
	
	\begin{table}[!t]
		\caption{Representation model and tool support for investigated notations.}
		\label{Poduct_ModelBased}
		\tiny
		\rowcolors{2}{gray!20}{white}
		\begin{tabularx}{4.5in}{ll|* 6{ X }c }
			\toprule
			
			\rowcolor{white}	
			\multicolumn{1}{c}{} &\multicolumn{1}{c}{} &
			\multicolumn{3}{c}{Final product} & \multicolumn{3}{c}{Supported Model}  \\\hline

			\rowcolor{white}
			\multicolumn{1}{l}{Notation}
			&\multicolumn{1}{l|}{Citation}                          
			& \rotatebox{70}{Tool}    
			& \rotatebox{70}{Prototype}   
			& \rotatebox{70}{None}   
			& \rotatebox{70}{UML based}    
			& \rotatebox{70}{DFD based	}   
			& \rotatebox{70}{Non UML/DFD }       \\
			\bottomrule
			\toprule
			ADM-RBAC & \cite{Diaz2008a}                                      &              	&\CheckmarkBold                &                &               &              &\CIRCLE                   \\
			Ahn-AC &\cite{Ahn2002}                                         &                 &               &\CheckmarkBold                 &\CIRCLE               &              &                   \\
			Alam-SECTET &	\cite{Ahn2002}                                  &                 &\CheckmarkBold               &                  &\CIRCLE               &              &                   \\
			AMF &\cite{Hu2010}                                           &                 &\CheckmarkBold               &                 &\CIRCLE               &              &                   \\
			Buyens-LP &	\cite{Buyens2013}                                     &                 &\CheckmarkBold               &                 &             &              &\CIRCLE                     \\
			FDAF &	\cite{Dai2007}                                          &                 &\CheckmarkBold               &                 &\CIRCLE               &              &                   \\
			Georg-AO &\cite{Georg2009}                                      &                 &               &\CheckmarkBold                 &\CIRCLE               &              &                   \\
			Giordano AC &\cite{Giordano2010a}                                   &                 &\CheckmarkBold               &                 &               &              &\CIRCLE                   \\
			Gomaa UML &	\cite{Shin2004}                                     &                 &               &\CheckmarkBold                 &\CIRCLE               &              &                   \\
			Hafner-SOA &\cite{Hafner-2005}                                    &                 &               &\CheckmarkBold                 &\CIRCLE               &              &                   \\
			Hoisl-SOA &	\cite{Hoisl2014}                                     &\CheckmarkBold                 &               &                 & \CIRCLE              &              &                   \\
			Kim-AC &	\cite{Kim2011}                                        &                 &\CheckmarkBold               &                 &\CIRCLE               &              &                   \\
			Kong-Threat &\cite{KONG2010}                                   &                 &               &\CheckmarkBold                 &              &              &\CIRCLE                    \\
			Mariscal-AC &\cite{Pavlich-Mariscal2010}                                   &                 &\CheckmarkBold               &                 &\CIRCLE               &              &                   \\
			Medina-DB& \cite{Trujillo2009}                                     &                 &\CheckmarkBold               &                 &\CIRCLE               &              &                   \\
			Memon-SECTET &	\cite{Memon2014}                                  &\CheckmarkBold                 &               &                 &\CIRCLE               &              &                   \\
			Nakamura-SOA &\cite{Satoh2006}                                  &                 &               &\CheckmarkBold                 &\CIRCLE               &              &                   \\
			PbSD &\cite{Abramov2012}                                          &                 &\CheckmarkBold               &                 &\CIRCLE               &              &                   \\
			Ray-AC &\cite{Ray2004}                                        &                 &               &\CheckmarkBold                 &\CIRCLE               &              &                   \\
			SecureSOA &	\cite{Menzel2010}                                     &                 &\CheckmarkBold               &                 &\CIRCLE               &              &                   \\
			SecureUML &	\cite{Basin2009}                                     &\CheckmarkBold                 &               &                 &\CIRCLE               &              &                   \\
			Sohr-AC &\cite{Sohr2005}                                       &                 &\CheckmarkBold               &                 &\CIRCLE               &              &                   \\
			UML AC &\cite{Koch2006}                                        &                 &               &\CheckmarkBold                 &\CIRCLE               &              &                   \\
			UMLS &	\cite{Heldal2003}                                         &                 &               &\CheckmarkBold                 &\CIRCLE               &              &                   \\
			UMLsec &\cite{Jurjens2008}                                        &\CheckmarkBold                 &               &                 &\CIRCLE               &              &                   \\
			Vela-DB-XML &\cite{Vela2012}                                   &                 &               &\CheckmarkBold                 &\CIRCLE               &              &                   \\
			Xu-Petri &\cite{Xu2006}                                      &                 &               &\CheckmarkBold                 &              &              & \CIRCLE                   \\
			Yu-AC &\cite{Yu2009}                                         &                 &               &\CheckmarkBold                 &\CIRCLE               &              &                   \\
			Hafner-SECTET &\cite{Hafner2006b}                                 &                 &               &\CheckmarkBold                 &\CIRCLE               &              &                   \\
			SPARTA &\cite{Sion2018}                                        &                 &\CheckmarkBold               &                 &               &\CIRCLE              &                   \\

			VandenBerghe-DFD & \cite{VanDenBerghe20173}							 &                 &               &\CheckmarkBold                 &               &\CIRCLE              &                   \\
			LINDDUN	 &\cite{Wuyts2015a}							 			 &                 &               &\CheckmarkBold                 &               &\CIRCLE              &                   \\
			Sion-LINDDUN	 &\cite{Sion2018b}							 	 &                 &               &\CheckmarkBold                &               &\CIRCLE              &                   \\
			Kaitiaki &	\cite{Liu2005}									 &                 &\CheckmarkBold               &                 &               &              &\CIRCLE                   \\
			Suzuki-UML &\cite{Suzuki1999}									 &                 &\CheckmarkBold               &                 &\CIRCLE               &              &                   \\
			VLDesk &\cite{Francese2004}										 &\CheckmarkBold                 &               &                 &\CIRCLE               &              &                   \\
			VESIG & \cite{diaz2010visual}							 			 &\CheckmarkBold                 &               &                 &               &              &\CIRCLE                   \\
			Superlog &\cite{Flake2000}							 		&                 &               &\CheckmarkBold                 &               &              &\CIRCLE                   \\
			VisPro &\cite{Zhang2001}							 			&\CheckmarkBold                 &               &                 &               &              &\CIRCLE                   \\
			FESA-UML &\cite{Farkas2009}							 		&                 &\CheckmarkBold               &                 &\CIRCLE               &              &                   \\
			Pounamu &\cite{NianpingZhu2005} 							 		&\CheckmarkBold                 &               &                 &               &              &\CIRCLE                   \\
			Milo &\cite{Rao2018}							 			&\CheckmarkBold                 &               &                 &               &              &\CIRCLE                   \\
			SiMoNa &\cite{DeMorais2018}							 			&                 &               &\CheckmarkBold                 &               &              &\CIRCLE                   \\
			Midgar &\cite{GonzalezGarcia2014}							 			&\CheckmarkBold                 &               &                 &               &              &\CIRCLE                   \\
			MetaEdit+ 5.5 &	\cite{Tolvanen2009} 							 	& \CheckmarkBold                &               &                 &               &              &\CIRCLE                   \\
			Microsoft Threat Modeling &	\cite{Microsoft}				    & \CheckmarkBold                &               &                 &               &\CIRCLE              &                   \\
			Secure*BPMN &	\cite{Cherdantseva2014}			    &                 &               &                 &               &              &                   \\
			
			\hline    	\midrule	
			\textbf{Total}&								 &12                 &16               & 18                &27               &5              &14                   \\  
			\bottomrule

		\end{tabularx}

	\end{table}

	\paragraph{\textbf{UML-based notations }} \label{UML-based}
	
	The majority of used notations to represent security concerns are founded to be a Unified Modeling Language (UML) based, as stated in \cite{VandenBerghe2017}. Since the main focus, in this paper, is non-functional requirements (NFRs), most of the notations that have been collected are concern about security. Consequently, as seen in Table \ref{Poduct_ModelBased} most of them found to be UML based (27 notations out of 47). Some of the design notations are created to only detect new vulnerabilities while others such as Georg-AO \cite{Georg2009} does not. In Georg-AO proposed notation \cite{Georg2009}, they assess if this attack cause a huge risk then some security mechanism will be applied to mitigate it.


	\paragraph{\textbf{DFD-based notations }}\label{DFD-based}
	As mentioned previously, most of the common security design notations are \textit{UML-based}. Data Flow Diagram (DFD) is another notation based that followed by some, such as VandenBerghe-DFD and SPARTA modeling notations \cite{VanDenBerghe20173}\cite{Sion2018}. In VandenBerghe-DFD, they proposed a model which is inspired by DFD with some security elements and attached with well-defined semantics. In this model, the expert developer knowledge's that used for identifying and mitigating the potential threats will be supported to guarantee the correctness of applied security solution. This model has been illustrated using a banking system to show how it can be used as a strong foundation for security by design paradigm.
	In SPARTA, they represented a prototype that simplifies the embedding of security and privacy. That's done by supporting the process of capturing security and privacy patterns in a DFD-based design then provide threat elicitation based on constructed knowledge.

	LINDDUN and Sion-LINDDUN both are another example that use DFD in their activities representation \cite{Wuyts2014} \cite{Wuyts2015a} \cite{Sion2018b}. LINDDUN is threat modelling methodology that is focusing mainly on privacy where a systemic approach for producing the privacy requirements is proposed. Identifying all potential privacy threats is done by iterating over the model elements. After that, the threats are manually assessed based on their importance (likelihood and impact). Likewise, Microsoft Threat Modeling, which follows a STRIDE, is threat modelling methodology for security.

	\paragraph{\textbf{Non UML or DFD based notations }}\label{non-UML-DFD}
	UML and DFD, as explained previously, are general-purpose modelling languages for many of the systems. On the other hand, some languages are Domain-Specific Modeling Languages (DSML) or very specialized ones (14 in total in this study) as seen in Table \ref{Poduct_ModelBased}. In DSML, a domain-specific language is used to represent a system such as SiMoNa \cite{DeMorais2018}. SiMoNa is an IoT Infographic Domain-Specific Modeling Language. It is built using MetaEdit+ 5.5 workbench software which uses a GOPPRR(Graph, Object, Port, Property, Relationship and Role) meta-modelling language. 
	
	Furthermore, there are very specialized visual language such as ADM-RBAC and Giordano-AC \cite{Diaz2008a} \cite{Giordano2010a}. ADM-RBAC (Ariadne Development Method with Role-Based Access Control) is used for modelling hypermedia and web systems area to specify RBAC access rules at integrated two abstraction levels. This notation is an extension of the Ariadne Development Method (ADM) which does not use UML for its representation. Instead, the visual model is designed for role-based access control (RBAC); where a role can be created with defined relations, assigned permissions and generated policies. In conceptual models, policies are specified using function specifications, authorization rules and the user diagrams. In detailed models, policies are specified using access tables in addition to some of the previously produced models from conceptual phase with more details \cite{Diaz2008a}. Likewise ADM-RBAC, Giordano-AC model \cite{Giordano2010a} is a visual model for role-based access control (RBAC) which does not use UML mainly for its representation. In Giordano AC system, there are many tools for enabling a different kind of users to edit the security policies visually. After that, the system will generate XACML (eXtensible Access Control Markup Language) code. 

	\subsection{Validation of the Notations}
	
	\begin{table}[!t]
		\caption{Validation techniques that used for each notation which can be Case studies, Experiments and illustrations. Novice, undergraduate, master, PhD and expert are the experiments’ participants educational background. (*)For Experiments column: the number represents the total number of participants in the experiment for each notation. Empty cell indicates there is no experiment done. (\textbf{-}) in SPARTA means that they stated there is an experiment, but they did not give details about the participants number or background.}
		\label{validation}	
		
		\tiny
		\centering
		\rowcolors{2}{white}{gray!20}
		\begin{tabularx}{4.5in}{ p{2.5cm}p{1.25cm}|| p{.5cm}p{.5cm}p{.5cm}p{.5cm}p{.5cm}  |   |p{.5cm}p{.5cm}p{.5cm} } 
			
			\toprule
			
			\multicolumn{1}{c}{}
			&\multicolumn{1}{c}{}
			& \multicolumn{5}{c}{\textbf{Participants}} 
			& \multicolumn{3}{c}{\textbf{Validation}}\\
			
			\rowcolor{white}
			\multicolumn{1}{c}{}
			&\multicolumn{1}{c}{}
			& \multicolumn{5}{c}{\textbf{background}} 
			& \multicolumn{3}{c}{\textbf{technique} }\\
			
			\bottomrule

			\multicolumn{1}{l}{\textbf{Notation}} 
			&\multicolumn{1}{l}{\textbf{Citation}}                         
			&\rotatebox{90}{Novice}    
			&\rotatebox{90}{Undergraduate}   
			&\rotatebox{90}{Master} 
			&\rotatebox{90}{PhD} 
			&\rotatebox{90}{Expert} 
			
			&\rotatebox{90}{Case studies}    
			&\rotatebox{90}{Experiments*}   
			&\rotatebox{90}{Illustrations}  \\ 
			\bottomrule
			\toprule


			ADM-RBAC& \cite{Diaz2008a}                           &        &                               & \CheckmarkBold &                               &                               &                               & 18  & \CheckmarkBold \\
			Ahn-AC &\cite{Ahn2002}                                &        &                               &                               &                               &                               &                               &                               &                               \\
			Alam-SECTET &	\cite{Ahn2002}                         &        &                               &                               &                               &                               &                               &                               & \CheckmarkBold \\
			AMF &\cite{Hu2010}                                  &        &                               &                               &                               &                               &                               &                               & \CheckmarkBold \\
			Buyens-LP &	\cite{Buyens2013}                            &        &                               &                               &                               &                               & \CheckmarkBold &                               & \CheckmarkBold \\
			FDAF &	\cite{Dai2007}                                 &        &                               &                               &                               &                               &                               &                               & \CheckmarkBold \\
			Georg-AO &\cite{Georg2009}                             &        &                               &                               &                               &                               &                               &                               & \CheckmarkBold \\
			Giordano AC &\cite{Giordano2010a}                        &        &                               & \CheckmarkBold & \CheckmarkBold & \CheckmarkBold & \CheckmarkBold & 20  &                               \\
			Gomaa UML &	\cite{Shin2004}                           &        &                               &                               &                               &                               &                               &                               & \CheckmarkBold \\
			Hafner-SOA &\cite{Hafner-2005}                            &        &                               &                               &                               &                               &                               &                               & \CheckmarkBold \\
			Hoisl-SOA &	\cite{Hoisl2014}                         &        &                               &                               &                               &                               &                               &                               & \CheckmarkBold \\
			Kim-AC &	\cite{Kim2011}                             &        &                               &                               &                               &                               & \CheckmarkBold &                               & \CheckmarkBold \\
			Kong-Threat &\cite{KONG2010}                           &        &                               &                               &                               &                               &                               &                               & \CheckmarkBold \\
			Mariscal-AC &\cite{Pavlich-Mariscal2010}                          &        &                               &                               &                               &                               &                               &                               & \CheckmarkBold \\
			Medina-DB& \cite{Trujillo2009}                           &        &                               &                               &                               &                               & \CheckmarkBold &                               & \CheckmarkBold \\
			Memon-SECTET &	\cite{Memon2014}                          &        &                               &                               &                               &                               &                               &                               & \CheckmarkBold \\
			Nakamura-SOA &\cite{Satoh2006}                         &        &                               &                               &                               &                               &                               &                               & \CheckmarkBold \\
			PbSD &\cite{Abramov2012}                               &        &  &  \CheckmarkBold                             &                               &                               &                               & 148 & \CheckmarkBold \\
			Ray-AC &\cite{Ray2004}                            &        &                               &                               &                               &                               &                               &                               & \CheckmarkBold \\
			SecureSOA &	\cite{Menzel2010}                            &        &                               &                               &                               &                               &                               &                               & \CheckmarkBold \\
			SecureUML &	\cite{Basin2009}                             &        &                               &                               &                               &                               &                               &                               & \CheckmarkBold \\
			Sohr-AC &\cite{Sohr2005}                             &        &                               &                               &                               &                               &                               &                               &                               \\
			UML AC &\cite{Koch2006}                                &        &                               &                               &                               &                               &                               &                               & \CheckmarkBold \\
			UMLS &	\cite{Heldal2003}                                  &        &                               &                               &                               &                               &                               &                               & \CheckmarkBold \\
			UMLsec &\cite{Jurjens2008}                              &        &                               &                               &                               &                               & \CheckmarkBold &                               & \CheckmarkBold \\
			Vela-DB-XML &\cite{Vela2012}                         &        &                               &                               &                               &                               & \CheckmarkBold &                               & \CheckmarkBold \\
			Xu-Petri &\cite{Xu2006}                              &        &                               &                               &                               &                               & \CheckmarkBold &                               & \CheckmarkBold \\
			Yu-AC &\cite{Yu2009}                              &        &                               &                               &                               &                               &                               &                               & \CheckmarkBold \\
			Hafner-SECTET &\cite{Hafner2006b}                       &        &                               &                               &                               &                               &                               &                               & \CheckmarkBold \\
			SPARTA &\cite{Sion2018}                        &                               &                               &                               &                               &                               &                               &\textbf{ -} &                               \\
			VandenBerghe-DFD & \cite{VanDenBerghe20173}                     &                               &                               &                               &                               &                               &                               &                               & \CheckmarkBold \\
			LINDDUN	 &\cite{Wuyts2015a}                              &                               &                               &                               &                               &                               &                               &                               & \CheckmarkBold \\
			Sion-LINDDUN	 &\cite{Sion2018b}                         &                               &                               &                               &                               &                               &                               &                               & \CheckmarkBold \\
			Kaitiaki &	\cite{Liu2005}                             &                               &                               &                               &                               &                               &                               &                               & \CheckmarkBold \\
			Suzuki-UML &\cite{Suzuki1999}                           &                               &                               &                               &                               &                               &                               &                               & \CheckmarkBold \\
			VLDesk &\cite{Francese2004}                               &                               &                               &                               &                               &                               & \CheckmarkBold &                               & \CheckmarkBold \\
			VESIG & \cite{diaz2010visual}                                &                               &                               & \CheckmarkBold &                               &                               &                               & 12 & \CheckmarkBold \\
			Superlog &\cite{Flake2000}                              &                               &                               &                               &                               &                               &                               &                               & \CheckmarkBold \\
			VisPro &\cite{Zhang2001}                                &                               &                               &                               &                               &                               & \CheckmarkBold &                               & \CheckmarkBold \\
			FESA-UML &\cite{Farkas2009}                              &                               &                               &                               &                               &                               &                               &                               & \CheckmarkBold \\
			Pounamu &\cite{NianpingZhu2005}                              &                               &                               &                               &                               &                               &                               &                               & \CheckmarkBold \\
			Milo &\cite{Rao2018}                                &                               & \CheckmarkBold &                               &                               &                               &                               &                   20            & \CheckmarkBold \\
			SiMoNa &\cite{DeMorais2018}                                &                               &                               &                               &                               &                               &                               &                               & \CheckmarkBold \\
			Midgar &\cite{GonzalezGarcia2014}                               & \CheckmarkBold &                               &                               &                               & \CheckmarkBold &                               & 21 & \CheckmarkBold \\
			MetaEdit+ 5.5 &	\cite{Tolvanen2009}                        &                               &                               &                               &                               &                               &                               &                               & \CheckmarkBold \\
			Microsoft Threat Modeling &	\cite{Microsoft}           &                               &                               &                               &                               &                               &                               &                               &        \CheckmarkBold                       \\
			Secure*BPMN &	\cite{Cherdantseva2014}                         &                               &                               &  \CheckmarkBold                             &                               &\CheckmarkBold                               &        \CheckmarkBold                       &          31                     &\CheckmarkBold      \\

			\hline    	\midrule
			\textbf{Total}	&				               &   1            &   1              &     5          &       1       &  3              &     10        &	270	 &   43     \\

			\bottomrule
			
		\end{tabularx}
	\end{table}

\begin{figure}
		\includegraphics[scale=.65]{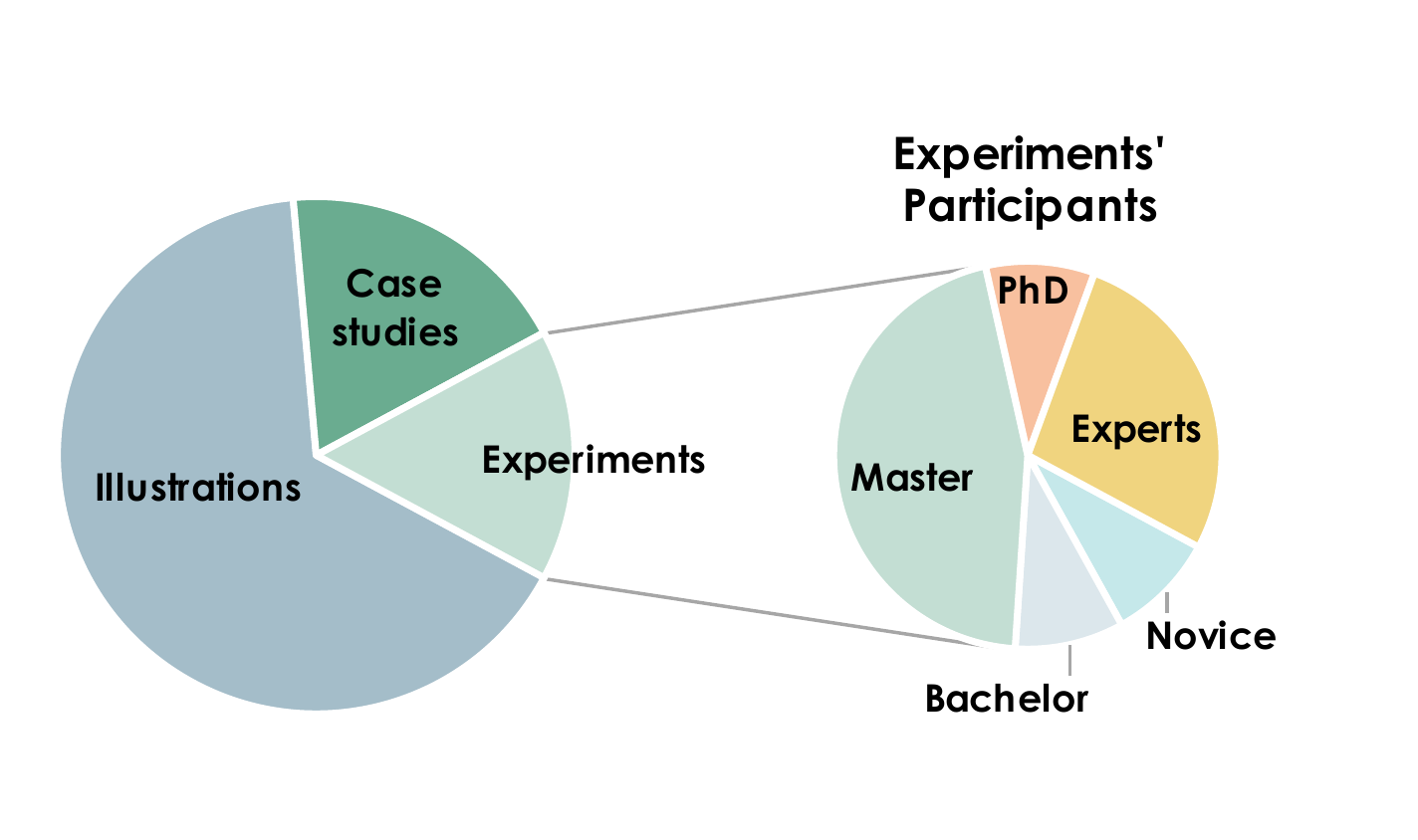}
		\caption{Pie chart (left) shows validation type. Pie chart (Right) shows experiment participants educational background.}
		\label{fig:participantsBG}
	\end{figure}

	As seen in Table \ref{validation}, a summary of the performed validation type that has been done for each design notation along with the number and educational background of participants if there is any. It is noticeable that the majority of the notations lack case studies and experiments; in contrast, most of them have been illustrated in one or multiple examples. However, most of the illustration examples are abstract and lack of deep details. Only a few of the notations used experiments (10 out of 47) and some have case studies  (10 out of 47) as a way of validation. In the experiments, it is noticeable that most of the experiments` participants have master degree as seen in Figure \ref{fig:participantsBG}. Besides that, some do not have any kind of illustration$ \ $validation such as Sohr-AC and Ahn-AC.
	
	While some notations such as Giordano AC uses a group of heterogeneous people such as from universities, industrial managers and technicians, others like ADM-RBAC and PbSD, uses a group master and bachelor students respectively. In ADM-RBAC \cite{Diaz2008a}, a comparison has been done on the experiment; where 18 evaluators are divided into two groups. One of the groups trained while the other is not. All the evaluators are first-year master students, and the evaluation is part of a Hypermedia Design course. These experiments are done on a small scope and lack of deep analysis. In Giordano AC \cite{Giordano2010a}, the authors validate their model using both the use case and experiment. The case study done by using 20 evaluators who are managers and technicians are from universities and industries, participated to test system usability using a technique called think.
	In the experiment, they compared their notation to the most related visual-based tool called XGrid. The evaluators participated for three days and follow the think-aloud technique using a one-to-one session. Each validator has an introductory course for 90 minutes about the two systems and their notations. After that, they were asked to use these systems for 20 minutes with tutor support. Then, they were asked to apply three different scenarios with using the systems tools in the definition of access policies. The scenarios include file hosting service, Massively multiplayer online role-playing game (MMORPG), and content management system. This experiment is done on a small scale (only 6 participants) and lacks an in-depth analysis.
	In the use case, they produce a Multimedia Content Management System (MCMS) as a collaboration with a software development company and embed access control policies using their approach. MCMS is based on open source content management system called OpenCMS, that provide extending some modules simply. By applying their approach, the authorizing and administrating users are simplified.

	In PbSD \cite{Abramov2012}, 148 third-year undergraduate students participated in the experiment from (Security of Computers and Communication Networks) course. The students are from different departments: Information Systems Engineering (ISE) and Software Engineering (SE) programs and were divided into two groups to apply some tasks to see the security requirements comparison between PbSD with SQL and Oracle’s VPD. 
	
	Likewise PbSD, VESIG  initial tool was tested in (Formal Methods for Web Design) course using 12 master students on Science and Technology in Computing \cite{diaz2010visual}. The required design task is done by pairs for 90 minutes divided as: 10 minutes was an introduction to the tool from VESIG team, 45 minutes to draw a sketch on a paper, and the remaining time is used to apply the sketch using the tool and completing 6 open questions. In the end, 6 different design and 12 questioners are collected.
	Not only postgraduate students are the one used for evaluation, but Milo web-based visual programming experiment is also done on 20 computer science undergraduate students \cite{Rao2018}.
	
	In Midgar \cite{GonzalezGarcia2014}, the experiment is done based on 21 participants who tested individually. 12 of the participants are software developers and 9 of them are people who interested in IoT. They have been asked to evaluate the applications' editor and generator from the developer and the user point of view. In SPARTA \cite{Sion2018}, they stated that the evaluation is done by running a risk analysis on the WebRTC, submitting SPARTA system to SecureDrop system, and comparing it with Microsoft Threat Analysis performance.

	Some of the notation their case studies are similar to an example scenario. For example, Kim-AC case study is presented as an application for online shopping while Xu-Petri case study is about modelling real-world shopping cart application \cite{Xu2006} \cite{Xu2006}. Both of them is applying STRIDE technique to identify threats. There are no participants for real and extensive analysis.  Medina-DB on the other hand, has a deep case study about the Secure Data Warehouse for several pharmacies done by the authors themselves in separate paper \cite{soler2007application}. Likewise, VisPro and UMLsec have some examples and case studies some of them have some details \cite{Jurjens2008}\cite{Zhang2001}.

	\section{Design Principles (Tooling) for Non Functional Requirements}\label{Section:DesignPrinciples}
	So far reviewed number or different design notation and languages that are being developed to design application. The next most important step is to make these notations available for the software engineers to use within SDLC. For example, Visual Paradigm (visual-paradigm.com) is a  Tool  that has been developed to make design notations and languages (such as UML 2, SysML and Business Process Modeling) available for software engineers to use. These tools are designed to augment the software design capabilities of software engineers and to make the software development process efficient and effective (e.g., help to designs faster, help to reduce human errors/mistakes). In this section, we are going to review different design principles and how they are being applied in existing design tools (related to security / privacy / IoT).

	Designing is considered as an easy thing to do when everything goes smoothly and right until a mistake or misunderstanding happened then complications occur rapidly. One of the solutions is adapting a human-centred design (HCD) approach \cite{Jun2008}. In this approach human needs, behaviours and capabilities are taken into consideration in the first place. Therefore, understanding psychology should go with understanding the technology side by side. Additionally, this approach adapts building good communication to achieve good design especially if there is a machine required in the design. This is done for understanding the interaction between machines and humans. It is more important for designers to understand what will happen if things go wrong more than what will happen if it goes as planned.

	
	\subsection{Fundamental Principles of Interaction} 
	There are many principles of good interaction design and one of them is Norman \cite{norman2013design}. He divides the principles to five which are: affordances, signifiers, mapping, feedback, and conceptual model. Three tools will be evaluated based on these principles later to see who it can be applied. It can be seen in Table \ref{CompDesignPrinciples} which are: ViSiT, Microsoft threat modelling and ThreatDragon.

	\begin{table}
		\footnotesize
		\caption{Comparison between three tools: ViSiT, Microsoft threat modeling and Threat Dragon in term of principles of good interaction design.(*) ViSiT is analyzed based on the Figures on paper. (**) Not mentioned on the paper.}
		\label{CompDesignPrinciples}
		\begin{tabular}{|p{1.5cm}|p{3.5cm}|p{3.5cm}|p{3.5cm}|}
			\toprule
			& {\small ViSiT* }     & {\small Microsoft threat modeling} & {\small ThreatDragon} \\                                                                                                                                                                                                                                                                                                                      
			\toprule
			
			{\scriptsize AFFORDANCES}
			& It has visible affordance where user can understand the idea of dragging and dropping expressions then running the transformations.                               
			& It has visible affordance where user can understand the idea of drawing DFD as any other DFD drawing tool. Then it can do the threat analysis. It also supports creating custom threat templates.
			&   It has visible affordance where user can understand the idea of drawing DFD. However, it does not support drag and drop which is used to be in most of designing tools.                                                                                                                                      \\
			\hline
			
			SIGNIFIERS
			& It supports many signifiers such as: \begin{itemize}
				\item icons (zoom in/out, help, delete expression)
				\item buttons (save, run ..etc)
				\item gridline
			\end{itemize} 
			
			\includegraphics[scale=0.18]{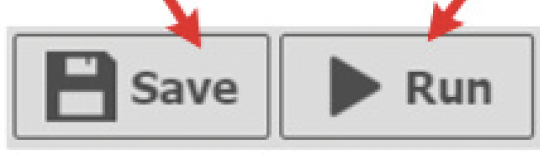}
			
			\includegraphics[scale=0.2]{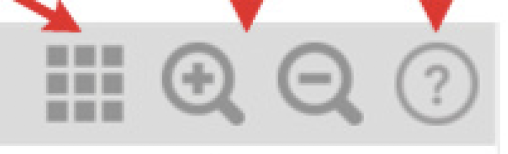}

			& It supports many signifiers such as:\begin{itemize}
				\item icons (zoom in/out, help)
				\item buttons (save, run ..etc)
				\item drop-down menus
				\item grid-line
			\end{itemize}However, linking two objects is not clear due to lacking for signifiers to say how to do that. It takes many tries to understand that one of the ways is selecting the two object then right-click to choose the linking option. 
			&It supports many signifiers such as:\begin{itemize}
				\item icons (zoom in/out)
				\item drop-down menus
			\end{itemize}However, the design is simple and lack of extra buttons and more signifiers.

			\includegraphics[scale=0.3]{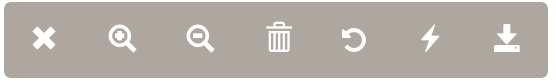}
			
			The green arrow above the object:

			\includegraphics[scale=0.42]{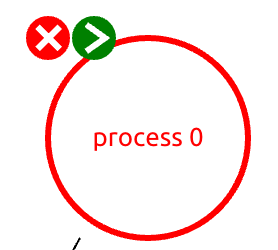}
			\\
			\hline
			
			MAPPING 
			&   When an object is taken from the target and properties collection, the object automatically placed in the shown grid at the left-hand side.
			It supports drag and drop from the toolbox.
			
			&   Mapping is supported.                                                                                                                                                                                                                                                                                                                                                 
			&Low mapping support. When the object from the toolbox is double-clicked, it appears in the canvas.  Having toolbox located at the left contains many objects, sometimes indicates dragging the objects to the canvas; however, it does not support drag and drop which is used to be in most of designing tools.                                                       
			\\
			\hline
			FEEDBACK  
			&     N/A**                                                                                                                                                              
			&  Feedback is supported. For example, when a developer is changing the name of any object from the stencils, it appears immediately at the object in the main canvas.        
			&There is feedback for example when the arrow is pressed message appears to tell the user what to do. However, the response time is slow, and the user can doubt about the request will be done or not.

			The message:
			
			\includegraphics[scale=0.15]{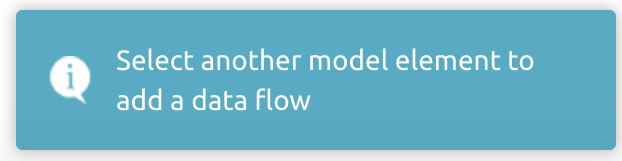}
			\\    \bottomrule                                                                                                                                                                                                                                                                                                                                               
			
		\end{tabular}
	\end{table}
	

	Affordance term means the relationship between the individual and object, in the way of how the person can understand the features of the object and how it can be used \cite{norman2013design}. The physical object should express how people can interact with them and this should be visible in the design. For designers, affordance distinguishability is very critical because visible affordances deliver clear signs to the operations of objects. Users should not pay a lot of effort to understand the proper way of how to use the features of things.

\begin{SCfigure}
		\includegraphics[scale=0.45]{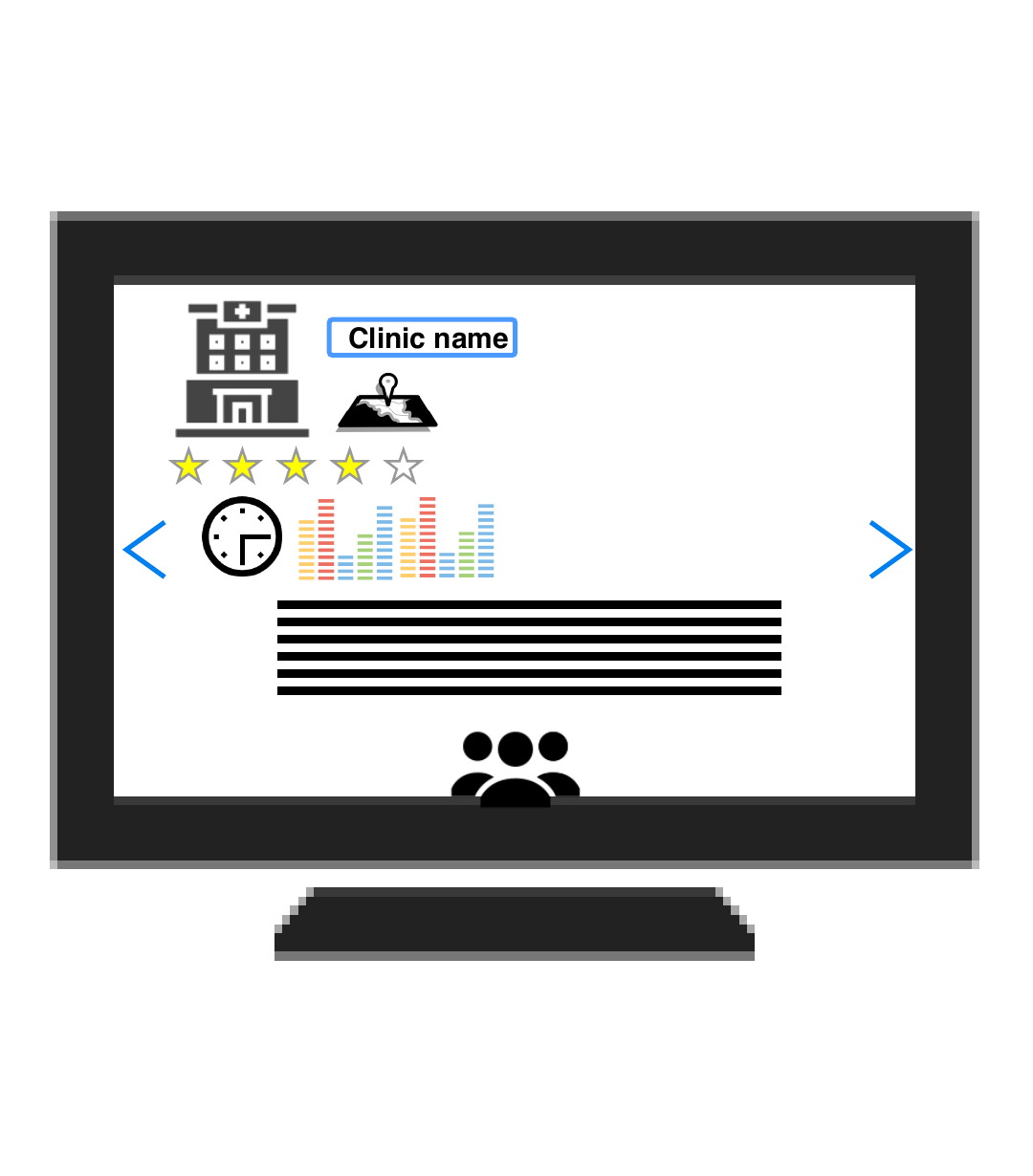}		
		\caption{Icons and arrows signifiers are used here on a touch screen. They offer
			signals about the produced actions for the clinic.
			Swiping left and right arrows move to new clinic suggestions. Swiping down, to see people recommendations and reviews. Clicking the map pin icon will direct to the clinic location on the map; clicking on the clock will show the opening hours.}
		\label{SignifiersTouch}
		\vspace{-30pt}
	\end{SCfigure}

	Signifiers are the guidance to the user of how to follow what the object can afford and do it in the right way. Designers have the responsibilities to design something understandable and quickly picked by the users. If there is a touch screen where the users need to touch something in order to go to other windows, there should be signifiers to tell them where exactly to touch. For example, there is a clinic profile as seen in Figure     \ref{SignifiersTouch}. This page has many icons and arrows which offer signals about the produced actions for the clinic. Swiping up left and right arrows move to new clinic suggestions. Swiping down is a sign to see people recommendations and reviews. Clicking the map pin icon will direct to the clinic location on the map.           Affordance and signifiers are easy and common to mix in between         \cite{norman2013design}, so in this case, touching is what the object affords, the arrow is the sign where the user should apply the touch.
	If we apply this principle to ViSiT \cite{2015iv}, the tool UI has many signifiers such as:     Having grid-lines in the canvas that guide the user to place the target collections and properties, having many buttons such as the one to save and run the transformation, having icons to zoom canvas or ask for help.

	Mapping in simple words is the relationship between two things. In design, it is easy to use object if there is a mapping between the layout and device. For example, if there is an arrow pointing up at the touch screen, it supposed when it is pressed to move the object up not down, left or right. A good design put in mind, during planning, how people behave to develop mapping.  IoT design application tool should consider this principle while implementing the UI. For Example, in Table \ref{CompDesignPrinciples} ViSiT \cite{2015iv}  has the icon of a floppy disk which is known as mapping for saving the task.
	
	Feedback: is away of allowing the users to know that the request is under process. It is a kind of communication with the users, so they do not feel ignored or lost. Good design must balance between ignoring the feedbacks or overusing them which both are annoying and lead to stop using the object (such as system software). So, while designing a designing tool this should be kept in mind to balance between asking and annoying users about every single privacy concerns, and keeping them notified about the critical one and keep some of them as default settings.
	
	
	A conceptual model as it is defined in some sources as “a high-level description of how a system is organized and operates'' \cite{johnson2002conceptual}.     In general, it is a simplified and easy way to explain to people about how something going to work. For example, using folders and icons in a computer OS is a way of helping users to create the conceptual model; while in the reality this is not the way files are organized and saved inside the computer.     There are many different conceptual models such as the one shown in the product technical support; however, the one that most concerned in designing is the mental model \cite{Jun2008}.      The mental model reflects how people understand how things work. The same item could have different mental models created by different people. In fact, a person may have a variety of mental models for the same item depending on his interpretation for each part's function.    The good conceptual model should be clear and understandable by the support of using affordances, signifiers and constraints. 
	
	By adopting a good conceptual model in IoT application designing tool, it will be easier for the developer to predict the effects and what plan to follow if things go wrong and not as planned while designing a data flow for an IoT application. If the model is complex that means the user should read the manual, again and again, to understand how things work. 
	
\begin{figure}
		\centering
		\includegraphics[scale=0.35]{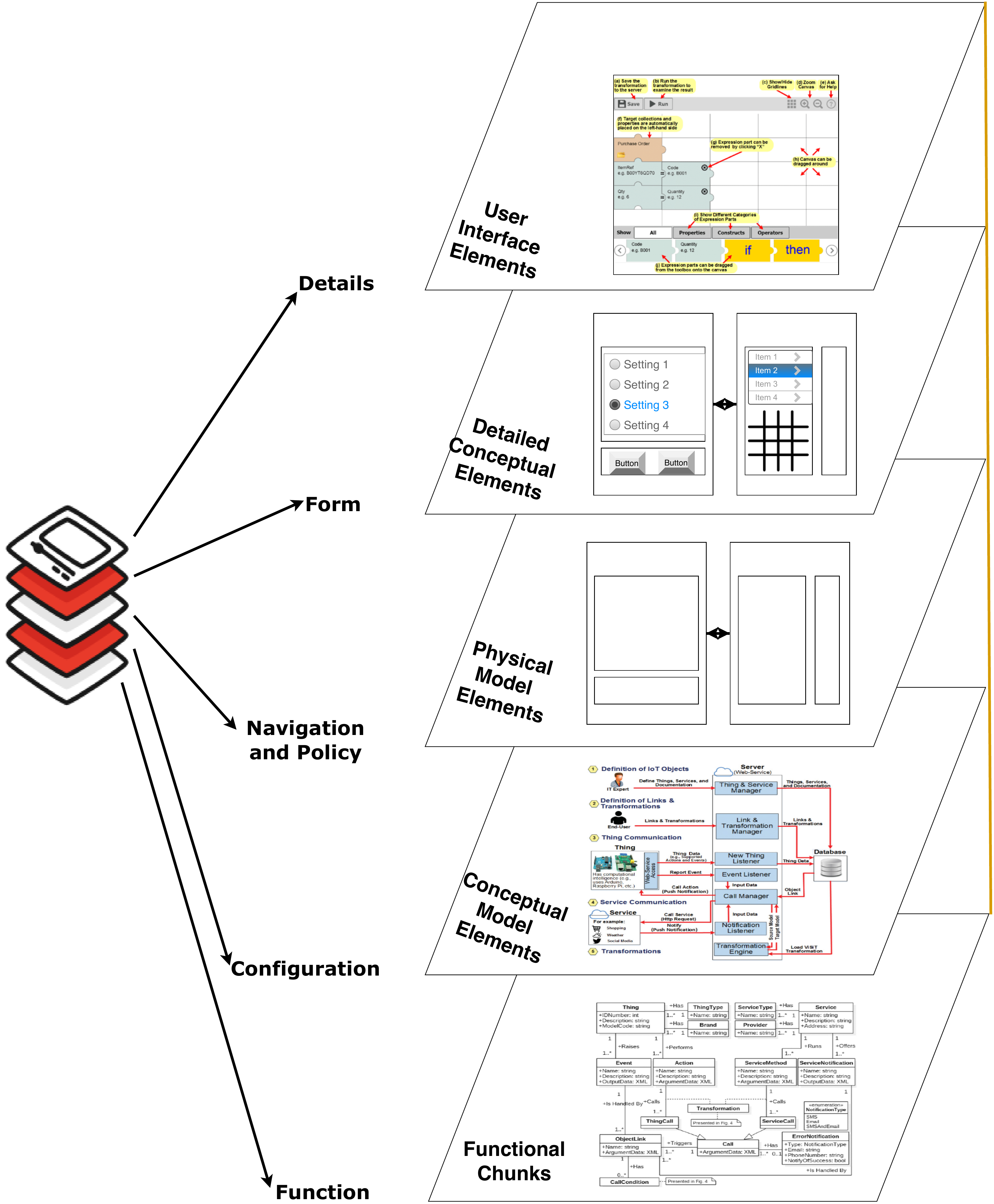}
		\caption{Parush layered framework \cite{2015iv} that represents conceptual model for interactive systems is applied on ViSiT IoT tool \cite{Akiki2017} to explaine how it can adapted while developing an interactive IoT designing tool.}
		\label{ConeptualViSiT}
	\end{figure}

	Parush \cite{2015iv} adds some details and divides the conceptual model for interactive systems it into five layers framework. These layers starting from bottom up are: function, configuration, navigation and policy, form and details layer. For illustration, all these layers will be mapped to Visual Simple Transformations tool called ViSiT \cite{Akiki2017} as seen in Figure \ref{ConeptualViSiT}. This tool is enabling the end-user to connect the Internet of Things (IoT) entities (services and things). It visualizes IoT solutions by using the metaphor of the jigsaw puzzle for specifying the transformation. This hierarchical conceptual model could be adapted while developing an interactive IoT designing tool.

	In the bottom is the function layer where a group of objects and tasks and their properties are stated to fulfil the required work such as how thing and service in ViSiT tool interact with each other. Configuration layer represents the abstraction of the functions and its relationship between its elements. It can be defined as functional architecture where the configuration of the conceptual model elements is based on the elements' relationship. In websites design, functional and information architecture are equivalent but in conceptual design is not. Information here is one of the elements as well as the object and its parameters. The navigation and policy layer describes the interactive system where navigating takes the user from place to another place based on predefined navigation map for routing instructions'. For example, if the user wants to drag a puzzle piece into the canvas (physical place) will it be in the same window (place) or dragged into other windows (places). Form layer is the layer that transforms from a high abstraction level to more detailed one that can be designed later into a user interface (UI). In this layer, many decisions are made which could affect the performance and usability of the system later on. The details of function chunks are mapped into physical places and connected with some actions such as where to fill in the properties and values of one event of ViSiT in order then to link it with a method. Finally, the user interface is the top layer of the conceptual model where all the details of graphical representation and visual layout are specified such as locations, colours, fonts and icons of the items.

	%
	%
	%

	As said  ViSiT IoT tool supports somehow detailed model while other tools do not. Threat Dragon, which is threat modelling tool, is more basic without any clear support to IoT. The threat is represented simply without further details in the right side of the tool windows in the form of chick boxes. On the other hand, the Microsoft threat modelling tool supports IoT security threats in the last version (2018). In Microsoft, after pressing analyze button the threats are presented in the bottom side of the same windows. All the threat is classified based on the term of threat title, category, description and priority...etc. Some of them even linked to the Microsoft website for further explanation about the threat and how it can be overcome. A short comparison between both of them (Microsoft threat modelling and Threat dragon) is illustrated in Table \ref{ThreatDragonVsMTM}.
	
	As will as ViSiT IoT tool, some analyzed notations such as ADM-RBAC \cite{Diaz2008a} supports detailed models for access specification to provide further details from the previous phase. In contrast, both Microsoft Threat Modeling tool and Threat Dragon tools have a single level of view where everything is shown in the same window and they do not support any kind of multiple detailed views or moving into other windows. Both approaches, basic or detailed user interface can be mixed to have an IoT designing tool that affirms to what user needs.

	\begin{table}[]
		\footnotesize 
		\centering
		\caption{Threat Dragon compared to Microsoft threat modeling.}
		\label{ThreatDragonVsMTM}
		\begin{tabular}{l|  lllllll}
			& Web-based & Desktop-based & Open source & Representation & IoT support   \\
			\toprule
			Threat Dragon             				&   \CheckmarkBold    &      \CheckmarkBold      & \CheckmarkBold      &     DFD    &  \xmark             \\
			Microsoft threat modeling 		&   \xmark        &   \CheckmarkBold            &      \xmark       &     DFD           &   \CheckmarkBold          
			\\	\bottomrule 
		\end{tabular}
	\end{table}
	


	\section{Research Challenges  and Opportunities}\label{Section:ResearchChallenges}

	As explained previously IoT application design has some challenges and adding some changes or enhancements to the traditional SDLC are required.  Below are some of the presented solutions or enhancements to make IoT development more controllable such as integrating some threat modelling techniques during the design stage such as STRIDE, introducing a framework to support privacy, or enhancing the SDLC to include different views such as prosumerization.
	\subsection{Lack on Notations and Languages for Privacy}\label{sectools}


	
	One of the solutions to make a system secure while designing is using threat modelling techniques such as OWASP \cite{OWASP2018} and Microsoft Threat Modeling Tool \cite{Microsoft} during the design phase. 
	There are two main methodologies, STRIDE and DREAD, that have been used for most of threat modelling techniques \cite{Fernandes2018}. STRIDE, which Microsoft Threat Modeling Tool uses, is acronym made up for mapping the threat categories which are: Spoofing, Tampering, Repudiation, Information Disclosure, Denial of Service and Elevation of Privilege. DREAD acronym for risk calculation metrics which are: Damage Potential, Reproducibility, Exploitability, Affected Users and Discoverability. The difference between the two is that STRIDE is used for threat identification while DREAD is used for risk calculation. Both OWASP \cite{OWASP2018} and Microsoft Threat Modeling Tool \cite{Microsoft} are used for security threats and none of them used for privacy in particular.

	
	\subsection{Lack of Tools to Supplement Methods }

	\begin{table}[]
		\caption{Security and privacy concerns with its corresponding threats mapped with DFD
			elements that vulnerable to threats. (DF: Data flow, DS: Data store, P: Process, E: External
			entity). STRIDE and LINDDUN are proposed by the Security Development Lifecycle (SDL) \cite{howard2006security} and Privacy Threats in Software Architectures \cite{Wuyts2015a} respectively.
		}
		\label{table:LINDUUNNvsSTRIDE}	
		\footnotesize  
		\begin{tabular}{l|l|  llllll}
			\multicolumn{2}{l}{ 
			}    & Property                            & Threat                           & DF & DS & P & E \\
			\midrule
			\cellcolor[HTML]{DAE8FC}                          & 	\cellcolor[HTML]{DAE8FC}                             & Authentication                      & \textbf{S}poofing                          &    &    & \textbullet & \textbullet \\
			\cellcolor[HTML]{DAE8FC}                          &   	\cellcolor[HTML]{DAE8FC}                           & Integrity                           & \textbf{T}ampering                         & \textbullet  & \textbullet  & \textbullet &   \\
			\cellcolor[HTML]{DAE8FC}                          &     	\cellcolor[HTML]{DAE8FC}                         & Non-repudiation                     & \textbf{R}epudiation                       &    & \textbullet  & \textbullet & \textbullet \\
			\cellcolor[HTML]{DAE8FC}                          &     	\cellcolor[HTML]{DAE8FC}                         & Confidentiality                     & \textbf{I}nformation Disclosure            & \textbullet  & \textbullet  & \textbullet &   \\
			\cellcolor[HTML]{DAE8FC}                          &    	\cellcolor[HTML]{DAE8FC}                          & Availability                        & \textbf{D}enial of Service                 & \textbullet  & \textbullet  & \textbullet &   \\
			\multirow{-6}{*}{\cellcolor[HTML]{DAE8FC}\rotatebox{90}{Security}}  &	\cellcolor[HTML]{DAE8FC}   \multirow{-6}{*}{\rotatebox{90}{STRIDE}} & Authorization                       & \textbf{El}evation of Privilege            &    &    & \textbullet &   \\

			\cellcolor[HTML]{EFEFEF}                          &  	\cellcolor[HTML]{EFEFEF}                           & Unlinkability                       & \textbf{L}inkability                       & \textbullet  & \textbullet  & \textbullet & \textbullet \\
			\cellcolor[HTML]{EFEFEF}                          &   	\cellcolor[HTML]{EFEFEF}                          & Anonymity and pseudonymity          & \textbf{I}dentifiability                   & \textbullet  & \textbullet  & \textbullet & \textbullet \\
			\cellcolor[HTML]{EFEFEF}                          &     	\cellcolor[HTML]{EFEFEF}                        & Plausible deniability               & \textbf{N}on-repudiation                   &    & \textbullet  & \textbullet & \textbullet \\
			\cellcolor[HTML]{EFEFEF}                          &    	\cellcolor[HTML]{EFEFEF}                         & Undetectability and unobservability & \textbf{D}etectability                     &    & \textbullet  & \textbullet & \textbullet \\
			\cellcolor[HTML]{EFEFEF}                          &    	\cellcolor[HTML]{EFEFEF}                         & Confidentiality                     & \textbf{D}isclosure of information         &    & \textbullet  & \textbullet & \textbullet \\
			\cellcolor[HTML]{EFEFEF}                          &   	\cellcolor[HTML]{EFEFEF}                          & Content awareness                   & Content \textbf{U}nawareness               & \textbullet  &    &   &   \\
			\multirow{-7}{*}{\cellcolor[HTML]{EFEFEF}\rotatebox{90}{Privacy}} & 	\cellcolor[HTML]{EFEFEF} \multirow{-7}{*}{\rotatebox{90}{LINDDUN}}  & Policy and consent compliance       & Policy and Consent \textbf{N}on-compliance &    & \textbullet  & \textbullet & \textbullet
		\end{tabular}
		\vspace*{-5mm}
	\end{table}

	\begin{figure}
		\centering
		\includegraphics[scale=0.25, width=13.5 cm]{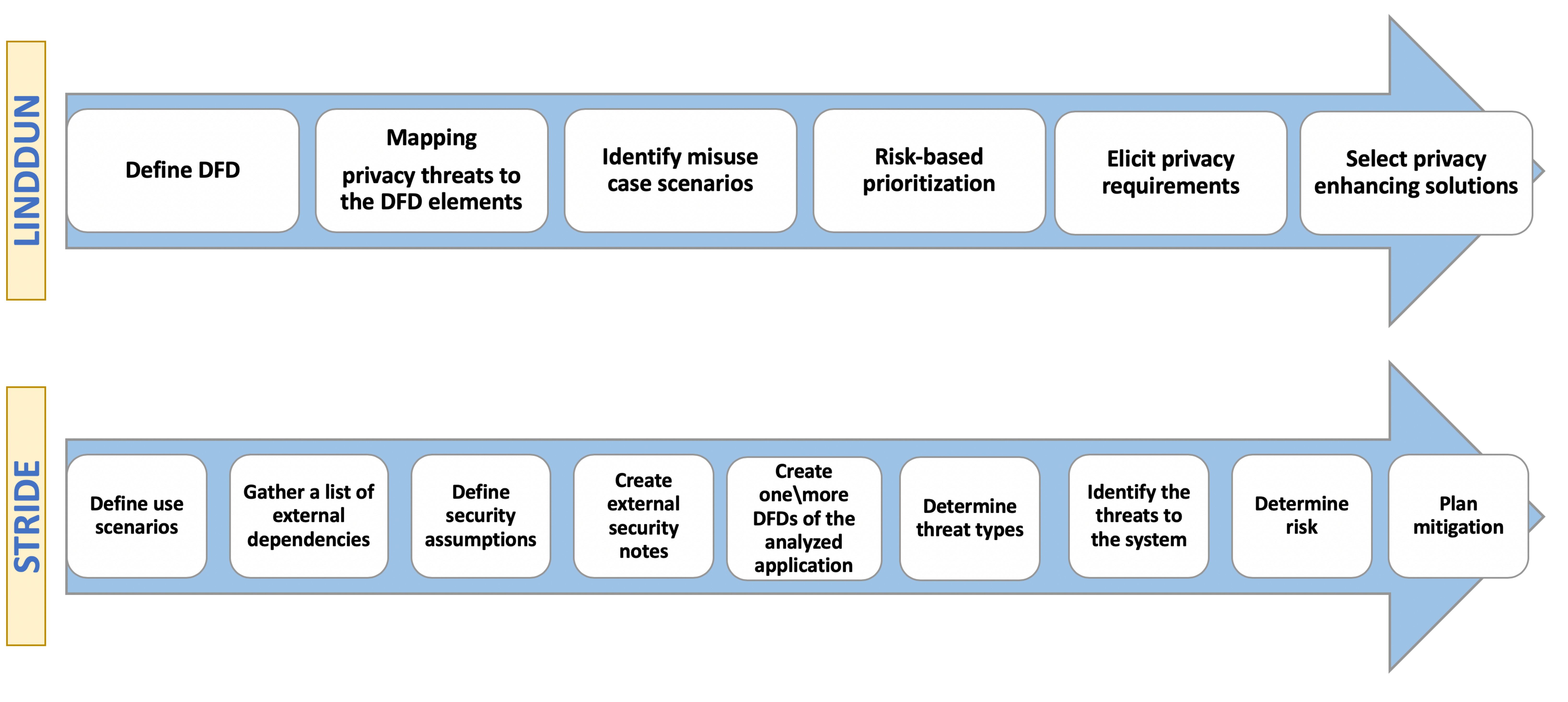}
		\caption{LINDDUN vs STRIDE Methodology}
		\label{LinddunVsStride}
		\vspace*{-5mm}
	\end{figure}

	As stated in Section \ref{sectools}, there are some tools that are used for security threats with absence for a framework for privacy. Since the main focus of STRIDE is security threat modelling, therefore LINDDUN was introduced to cover the privacy requirements. LINDDUN is a model-based technique$/$ framework that has been done to  focus exclusively on modelling privacy threats in software-based systems \cite{Wuyts2015a}. It is an acronym for seven types of threats: Linkability, Identifiability, Non-repudiation, Detectability, information Disclosure, content Unawareness and policy, and consent Non-compliance. All of these threats are mapped to the application DFD parts after it is created as seen in Table \ref{table:LINDUUNNvsSTRIDE}.

	STRIDE and LINDUDUNN have a different methodology in the way of the order of the steps that deal with the threats as seen in Figure \ref{LinddunVsStride}. In STRIDE, it is started with defining use case scenarios while it is the thirds step in LINDDUN which start with defining the DFD. As well as STRIDE, which adapted in Microsoft threat modelling tool to cover security concerns, we believe LINDDUN could be used to fulfil the privacy gap on other IoT tools. As stated, LINNDU was initiated as a framework but it does not have privacy pattern or tool yet.
	

	\subsection{Proactive Assistance}
	Proactive service means anticipating/predicting users’ concerns and addressing them proactively. For instance, most of the users when they have any concerns while browsing a website or using a tool, they will go to help or contact us option. Then, it becomes a waiting issue to resolve these concerns or questions. While in proactive service, the user concerns are predicted and solutions are offered based on the behaviour and the action have been made.
	
	By identifying and addressing any issue before it becomes a problem, availability of equipment and applications increases. Some places, such as hospitals, require high availability of critical hardware. For instance, Philips \cite{Philips} has a solution called e-Alert which is used to ensure high performance of magnetic resonance imaging (MRI) systems. It is done by sensing and monitoring the system continuously and responding quickly to any possible issues with MRI by issuing mobile messaging alerts that sent to biomeds and Philips service engineers. 
	Currently, many of the worldwide top-technical companies are tending to move from reactive to proactive engagement process. By supporting such a method, it will be possible to prevent any concerns before they arise. For example, Oracle \cite{Oracle} has proactive diagnostic support tools that predict system behaviour, prioritize actions, and prevent potential problems. 
	
	In IoT, the balance between the business's requirements and people's privacy is a big issue. Since IoT technology does not have completely standardize security and privacy requirements \cite{chaudhuri2018proactive}, there is a need for a proactive framework and assistant tool to help while developing IoT applications.
	There are tools with minimum proactive support or without proactive support especially the one that used during the designing phase of IoT applications.
	For example, if a developer wants to draw a data flow for a supermarket security system. There will be a smart camera in front of a supermarket which will be collecting Personally Identifiable Information (PII) such as human faces. The engineer does not know how long you keep the data, where it should be stored, and what is the compatible way to deal these data to agree with the Law.
	There is a need for a tool that waves the burden from the developer; whenever a camera is dragged and dropped,  privacy concerns will pop up automatically somewhere to give a hint about IoT privacy concerns. 
	We are focusing on the architecture/design part, where a data is flowing the privacy is preserved.

	\subsection{Integration of IoT Standards and Privacy
	}
	One of the studies that have been done is integrating OWASP standard Secure-SDLC with Microsoft Threat Modeling Tool in the IoT Application SDLC to produce a Secure-SDLC for IoT \cite{Fernandes2018}. In Fernandes \cite{Fernandes2018} study, privacy and security are demonstrated in healthcare application (IoT based Health Monitor). This application is used for tracking and monitoring the body temperature and pulse rate of hospitals’ patients. Here, security has been integrated while designing and implementing the system. The main idea is embedding security requirements from first to the final software$\backslash$system development stages, from requirements gathering to deployment. 
	
	At each stage, a number of known vulnerabilities are identified and mitigated. After that, testing for IoT based Health Monitor is made by using Open Web Application Security Project (OWASP) top 5 IoT vulnerabilities. 
	The SDLC is divided into five phases: Requirements Gathering, Design, Development, Testing and Deployment \cite{Fernandes2018}. Each phase has its level of own security review. In Requirement Gathering phase (planning and analysis), any security requirements and significant risks are identified and added at this stage. Security requirements can be associated with one of these: 
	authentication, authorization, error handling, session management, input validations, logging, secure communications, storage... etc.
	
	In the Design phase of the IoT solution, security threat modelling techniques are used for revising the design. At the development$\backslash$implementation phase, code revision is done to identify all coding errors that can cause any security risks. In the testing phase, the system is tested to confirm and validate that it meets the FRs as well as the NFRs. For example, testing the system using tests cases with OWASP Top 10 vulnerabilities. This step can make sure the system has a level of security. In the final stage, deployment, secure configurations are added to the system configurations with retesting for further variabilities \cite{Fernandes2018}.\\



	\subsection{Tools and Methods for Stakeholder Engagement through Codesign }
	
	Another way to develop IoT system is to follow End-user development (EUD) approaches for design and development such as the one represented in Martin work \cite{Martin2013}.
	End-user development (EUD) suggests that end-users could develop their own programs by giving them suitable tools that fulfil their requirements and save costs and efforts.
	
\begin{figure}
		\includegraphics[scale=0.35]{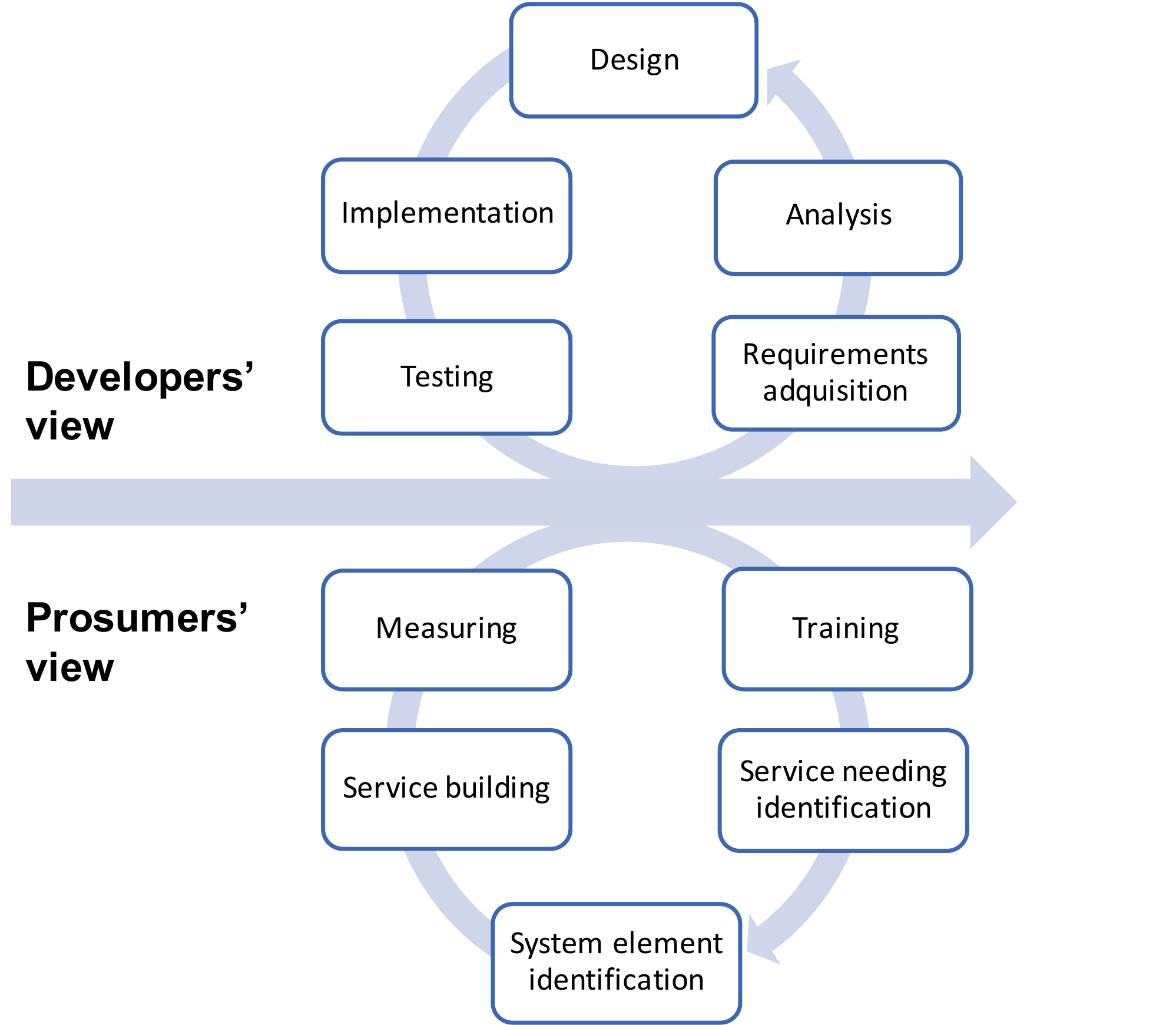}
		\caption{Methodological approach in Developers' and Prosumers' views based on \cite{Martin2013}. }
		\label{prosumer}
	\end{figure}

	In this Martin \cite{Martin2013}, a framework was created for enabling the prosumer users to create services in IoT scenarios since these scenarios are ideal for adopting the prosumer role. Prosumer is a combination of producer and consumer where the individual consumes and produces a product. The main goal for this framework is to enable the end-user (who does not have a high level of expertise) to be involved in development stages. The framework structured based on two interacted roles which are developers and prosumers. Each one of them has its own view as seen in Figure \ref{prosumer}.
	
	In the developers' view, the defined methodology has the five traditional stages of software development which are: requirement acquisition, analysis, design, implementation and testing. Each of these developer stages in the service creation framework should take into account the prosumer value.
	There are five more stages that present the prosumers' view which are: training, service needing identification system element identification, service building, and measuring. This framework was tested for creating and personalizing templates of web technologies in hospital pharmacy drug management. The built prototype was tested by allowing prosumer users, who have no technical expertise, such as pharmacy workers. They did some actions such as creating and sharing stock notifications for drugs available in the pharmacy.  This testing was made to make sure the presented approach has benefited from engaging the prosumer in the design phase to increase the usability of the system.
	
	\section{Conclusion}\label{Conclusion}
	There is evidence that number of IoT applications being developed are growing. This increasing is facing many complications in the field of designing and developing these applications such as supporting nodes heterogeneity, multiple hardware/software, multiple technologies, and developers' divers working together…etc. 
	With all of these complexities, all of the components are vulnerable to attacks. In IoT, security/privacy as non-functional requirements (NFRs) are critical due to its complicated nature; however, they tend to be ignored.
	
	This survey has stated the results of a literature review in the field of designing non-functional requirements in the internet of things. We examined, how such results can be applied to reduce IoT application design process complexity.
	In this survey, 47 notations, language, representation have been systematically scanned. Design principles, challenges and opportunities are also provided. The study has shown that most of the analyzed publications support security somehow, and rarely support privacy. It also highlights potential issues in supporting proactive design tools  for IoT privacy. Finally, we identified and discussed six research gaps related to privacy in IoT that need to be addressed.
	

	\bibliographystyle{ACM-Reference-Format}
	\bibliography{LReviewBibliography}
	
\end{document}